\documentclass{pasa}

\usepackage{graphicx}
\usepackage{todonotes}
\usepackage{animate}

\title[A gravitational wave follow-up strategy for ASKAP]{An optimised gravitational wave follow-up strategy with the Australian Square Kilometre Array Pathfinder}

\author[Dobie et al.]{D. Dobie$^{1,2}$\thanks{ddob1600@uni.sydney.edu.au}, T. Murphy$^{1}$, D.L. Kaplan$^{3}$, S. Ghosh$^{3}$, K.W. Bannister$^{2}$ and R.W. Hunstead$^{1}$
\affil{$^1$Sydney Institute for Astronomy, School of Physics, University of Sydney, NSW 2006, Australia}
\affil{$^2$ATNF, CSIRO Astronomy and Space Science, PO Box 76, Epping, NSW 1710, Australia}
\affil{$^3$Department of Physics, University of Wisconsin-Milwaukee, Milwaukee, WI 53201, USA}
}

\jid{PASA}
\doi{10.1017/pas.\the\year.xxx}
\jyear{\the\year}

\usepackage{aas_macros}

\usepackage{hyperref} 
\hypersetup{colorlinks,citecolor=blue,linkcolor=blue,urlcolor=blue}

\begin{document}

\begin{frontmatter}
\maketitle

\begin{abstract}
The detection of a neutron star merger by the Advanced Laser Interferometer Gravitational-Wave Observatory (LIGO) and Advanced Virgo gravitational wave detectors and the subsequent detection of an electromagnetic counterpart has opened a new era of transient astronomy. With upgrades to the Advanced LIGO and Advanced Virgo detectors and new detectors coming online in Japan and India, neutron star mergers will be detected at a higher rate in the future, starting with the O3 observing run which will begin in early 2019. The detection of electromagnetic emission from these mergers provides vital information about merger parameters and allows independent measurement of the Hubble constant. The Australian Square Kilometre Array Pathfinder (ASKAP) is expected to become fully operational early 2019 and its 30\,$\deg^2$ field of view will enable us to rapidly survey large areas of sky. In this work we explore prospects for detecting both prompt and long-term radio emission from neutron star mergers with ASKAP and determine an observing strategy that optimises the use of telescope time. We investigate different strategies to tile the sky with telescope pointings in order to detect radio counterparts with limited observing time, using 475 simulated gravitational wave events. Our results show a significant improvement in observing efficiency when compared with a na\"ive strategy of covering the entire localisation above some confidence threshold, even when achieving the same total probability covered.
\end{abstract}

\end{frontmatter}

\section{Introduction}
The detection of gravitational waves from a neutron star merger \citep[GW170817;][]{2017PhRvL.119p1101A} by the Advanced Laser Interferometer Gravitational-Wave Observatory (LIGO) and the subsequent detection of counterparts across the electromagnetic spectrum \citep{2017ApJ...848L..12A} heralds a new era in astronomy. A short gamma-ray burst \citep[sGRB;][]{2017ApJ...848L..13A,2017ApJ...848L..14G} was detected 1.7 seconds post-merger with thermal kilonova emission detected in the following hours \citep[][and references therein]{2017Sci...358.1556C,2017ApJ...848L..12A} and a non-thermal afterglow detected in X-rays \citep{2017Sci...358.1565E,2017ApJ...848L..25H,2017Natur.551...71T} and radio \citep{2017Sci...358.1579H} in the following days.

Observations over the first hundred days suggested that the observed X-ray and radio emission originate from the same source: optically-thin synchrotron emission \citep{2018Natur.554..207M}. The radio and X-ray light curves initially rose according to a power-law of the form $S_\nu(t,\nu) \propto t^{\delta_1} \nu^\alpha$ with spectral index $\alpha=-0.585\pm 0.005$ \citep{2018ApJ...856L..18M} and temporal index $\delta_1=0.84\pm 0.05$, peaked $149\pm 2$ days post-merger \citep{2018ApJ...858L..15D} and has declined according to a similar power-law with temporal index $\delta_2=-2.2\pm0.2$ or steeper \citep{2018ApJ...868L..11M,2018arXiv180806617V} out to 300 days post-merger.

These observations differ from the expected radio emission predicted prior to the event, with emission from a relativistic jet peaking within days of the merger and only detectable in a dense circum-merger medium, or late-time emission from sub-relativistic ejecta peaking on timescales of thousands of days \citep[e.g.][]{2012ApJ...746...48M,2016ApJ...831..190H}. Instead, this radio emission has been proposed to be produced by mildly relativistic quasi-spherical outflow \citep[a ``cocoon'';][]{2018MNRAS.473..576G,2018MNRAS.478..407N} with some contribution from an embedded off-axis structured jet \citep{2018PhRvL.120x1103L,2018ApJ...863L..18A,2018A&A...613L...1D,2018ApJ...856L..18M,2018ApJ...867...57R,2018Natur.561..355M}. We emphasise that we do not know whether the afterglow of GW170817 is typical of the population, although this will be answered as more events are detected.

Monitoring of the radio lightcurve allows us to estimate factors such as the circum-merger density, and energetics of the merger outflow, and may yet be able to determine the fate of the jet \citep{2018MNRAS.478..407N}. Combining Very Long Baseline Interferometry (VLBI) observations with monitoring of the radio lightcurve can also determine the inclination angle of the merger \citep{2018Natur.561..355M} and improve standard-siren constraints on $H_0$ \citep{2017Natur.551...85A,2018arXiv180610596H}

With upgrades to the Advanced LIGO and Advanced Virgo detectors, and the Kamioka Gravitational Wave Detector \citep[KAGRA;][]{2012CQGra..29l4007S,2013PhRvD..88d3007A} and LIGO-India \citep{2013IJMPD..2241010U} coming online, the detection rate of gravitational wave events will increase substantially to tens and possibly hundreds of events per year, localised to tens of $\deg^2$ \citep{2018LRR....21....3A}. Observation of a large number of events will enable better understanding of the neutron star merger population, and determine whether GW170817 was typical or exceptional. In particular, radio observations will constrain parameters including the isotropic equivalent energy and the circum-merger density and also the radio emission mechanism. 

In this paper we compare possible observing strategies for detecting radio emission from gravitational wave events with the Australian Square Kilometre Array Pathfinder \citep[ASKAP;][]{2008ExA....22..151J}. ASKAP is currently being commissioned and the final telescope will consist of 36 antennas, 12\,m in diameter, separated by baselines ranging from 37\,m to 6\,km resulting in angular scales of $10^{\prime\prime}$--$3^\prime$ at frequencies from 700--1800\,MHz. ASKAP is designed for all-sky surveys and uses MkII phased-array feeds \citep[PAFs,][]{6328742} consisting of a $6\times6$ array of beams, which produce a $30\,\deg^{2}$ field of view (FoV). Combined with its angular resolution and sensitivity, this makes ASKAP capable of following up poorly localised gravitational wave events with no other electromagnetic counterpart that would take hundreds of pointings to cover with other gigahertz-frequency radio telescopes. We analyse observing strategies previously designed for use with other telescopes, and determine the best way to implement them to optimise our use of ASKAP observing time.

\begin{table}
	\centering
	\caption{ASKAP design specifications and at the time of GW170817. Image RMS ($\sigma$) is calculated using the radiometer equation. Survey speed (SS) assumes a 100$\mu$Jy image RMS and ignores telescope overheads. The telescope is expected to reach design specifications in early 2019.}
	\label{tab:askap_specs}
	\begin{tabular}{lcc}
		\hline\hline
		 & for GW170817 & Design\\
		\hline
        Antennas & 16 & 36\\
		$T_{\rm sys}$ (K) & 70 & 50\\
		Bandwidth (MHz) & 240 & 300\\
        $\sigma_{\rm 10 min.}$ ($\mu$Jy/beam) & 300 & 82\\
        $\sigma_{\rm 12 hr.}$ ($\mu$Jy/beam) & 35 & 12\\
        SS ($\deg^2$hr$^{-1}$) & 21 & 270\\
        Frequency (GHz) & 0.7--1.4 & 0.7--1.8 \\
		\hline\hline
	\end{tabular}
\end{table}

\section{Searching for radio emission}
\subsection{ASKAP follow-up of GW170817}
We began follow-up observations of GW170817 with ASKAP 15 hours after the event, searching for coherent radio emission in fly's-eye mode \citep{GCN21562,GCN21671}. Imaging observations began two days after the event, covering the 90\% confidence region using the strategy presented in this paper \citep{GCN21625,GCN21639}; see \citet{2017PASA...34...69A} for further details.

\begin{figure}
	\includegraphics[width=\columnwidth,trim={0.5cm 0cm 0.1cm 0cm},clip]{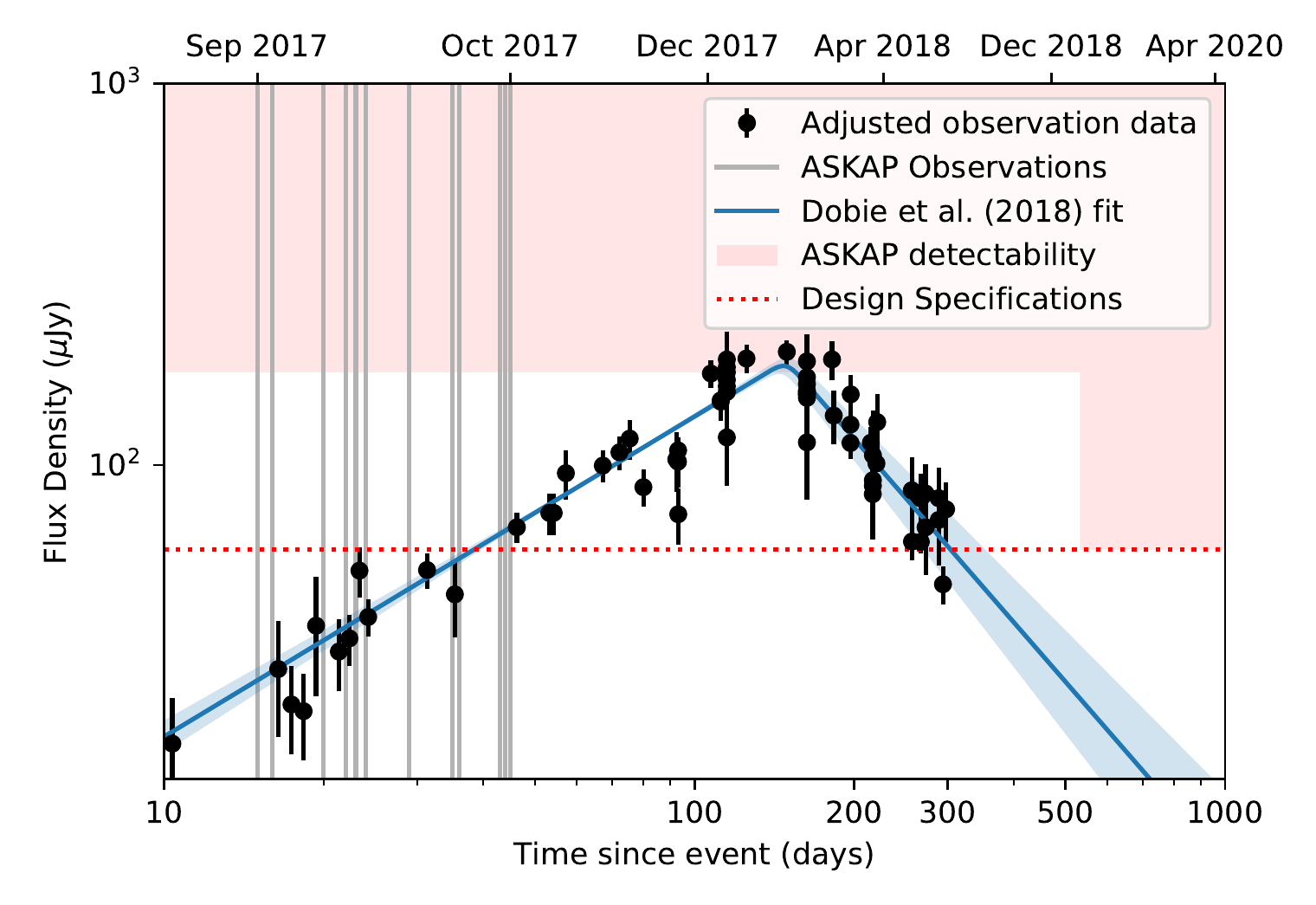}
	\caption{Detectability of GW170817 with ASKAP. Radio observations from \citet{2017Sci...358.1579H,2018Natur.554..207M,2018ApJ...858L..15D,2018ApJ...868L..11M} scaled to 1.4 GHz based on a spectral index of $\alpha=-0.58$ are noted in black, while the smoothed broken power law fit from \citet{2018ApJ...858L..15D} is shown in blue, with $1\sigma$ uncertainties shaded. The ASKAP $5\sigma$ detectability limits (see Table \ref{tab:askap_specs}) for current (shaded) and design (dotted line) specifications are shown. Vertical lines denote ASKAP observations of GW170817.}
	\label{fig:gw170817_prediction}
\end{figure}

Figure \ref{fig:gw170817_prediction} shows the lightcurve of GW170817 from \citet{2018ApJ...858L..15D} adjusted to 1.4\,GHz based on a spectral index of $\alpha=-0.58$, along with the $5\sigma$ detectability limit for the current telescope configuration and the design specifications (see Table \ref{tab:askap_specs}). The light curve peaked at the limit of ASKAP detectability and rapidly declined below this limit. However, if the telescope had been at full sensitivity for this event, it would have been detectable from 40 days post-merger and peaked at a flux density equivalent to a $15\sigma$ detection.

\subsection{Long-term synchrotron emission}
\label{sec:longterm_emission}
The focus of this paper is determining an observing strategy to detect the long-term radio emission (tens--thousands of days) from gravitational wave events, that optimises the use of telescope time. This emission could be produced, for example, by a relativistic jet launched after the merger \citep{2002ApJ...570L..61G} on timescales of days--weeks or the dynamical ejecta that caused the kilonova on timescales of months--years \citep{2011Natur.478...82N,2015MNRAS.450.1430H}.

The prompt gamma-ray emission from GW170817 is most likely driven by a jet, which produces a hot cocoon as it propagates into the merger ejecta \citep{2018MNRAS.473..576G}. As the cocoon expands quasi-spherically into the circum-merger medium, the high velocity components produce the early-time radio emission we have detected. A jet that successfully propagated through and breaks out of the cocoon (a structured jet) may drive the late-time radio emission. Continuum radio monitoring of GW170817 has been unable to distinguish between emission arising solely from a cocoon or structured jet \citep{2018MNRAS.478..407N}. VLBI observations suggest that the late-time radio emission is produced by a relativistic jet \citep{2018Natur.561..355M}, while polarisation measurements place constraints on the geometry and strength of the magnetic field within the jet \citep{2018ApJ...861L..10C}.

It may also be possible to detect long-term synchrotron emission where searches for kilonovae are limited by intrinsic factors like inclination angle, or extrinsic factors such as dust-obscuration, proximity to the Galactic Plane or solar angle. For example, optical and X-ray monitoring of GW170817 was limited by space telescope solar angle constraints, and the kilonova emission vital for localising the event would not have been detectable by ground based optical telescopes if the event had occured 3 months later in the daytime.

\subsection{Prompt radio emission}
It has been suggested that neutron star mergers may produce prompt, coherent radio emission. This could occur through magnetic braking as the magnetic fields of the two neutron stars synchronise \citep{2013PASJ...65L..12T}, an induced electromotive force in the final stage of the merger accelerating electrons to ultra-relativistic speeds \citep{2016ApJ...822L...7W}. The resultant supramassive neutron star may also collapse into a black hole \citep{2014ApJ...780L..21Z,2014A&A...562A.137F} and produce a relativistic plasma outflow \citep{2010Ap&SS.330...13P}. This emission would possibly be manifest as a signal similar to a Fast Radio Burst \citep[FRB,][]{2007Sci...318..777L,2013Sci...341...53T}, although there may be other forms of short-timescale radio emission present at the time of the merger such as pulsar-like behaviour \citep{1996A&A...312..937L}.

It may be possible to search for prompt radio emission from neutron star mergers with ASKAP, although this is not yet feasible due to observational constraints. During O2 the latency in receiving gravitational wave alerts was a few minutes and even after receiving the alert we are limited by observing overheads -- ASKAP can slew at a rate of $3\,\deg\textrm{s}^{-1}$ in azimuth and $1\,\deg\textrm{s}^{-1}$ in elevation. In contrast, the dispersion-induced time delay is expected to be $<10\,$s at 700 MHz \citep{2016MNRAS.459..121C}, the lowest frequency ASKAP band. However, the gravitational wave signal of the inspiral prior to GW170817 was detectable by Advanced LIGO $\sim 100\,$s pre-merger \citep{2017PhRvL.119p1101A}. In the future it may be possible to detect events prior to the merger \citep{2018PhRvD..97l3014C}, thereby allowing for telescopes to be on-source as the merger occurs.

ASKAP can observe in fly's-eye mode, where the telescope array is operated as a set of individual dishes pointed independently (and not necessarily contiguously). Each antenna has a FoV of $30\,\deg^2$, split into 36 beams in a $6\times6$ square, yielding a potential observing area of hundreds of $\deg^2$, allowing entire gravitational wave localisation regions to be covered instantaneously. Using the full 36 antenna array in fly's-eye mode allows for simultaneous coverage of 3\% of the sky. The probability of serendipitous detection of prompt radio emission can be increased by pointing the telescope at the region of sky where Advanced LIGO has maximum sensitivity. We note that due to the expected timescales for synchrotron radio emission discussed in Section \ref{sec:longterm_emission}, it is unlikely that low-latency observations in continuum mode will make any detection.

ASKAP has proven effective in detecting FRBs in fly's-eye mode, with 23 detected so far \citep{2017ApJ...841L..12B,2019ApJ...872L..19M,shannon2018dispersion}, making up over a quarter of all FRBs published to-date\footnote{\url{http://www.frbcat.org/} \citep{2016PASA...33...45P}.} It is capable of arcminute-localisation of coherent radio emission in fly's-eye mode, which is insufficient to positively identify the host galaxy but small enough for follow-up with other telescopes \citep{2018ApJ...867L..10M}. 

The detection of any form of coherent radio emission, coincident with a gravitational wave event would aid in localising and constraining the event in a similar way to the concurrent detection of GW170817 and GRB170817A \citep{2017ApJ...848L..13A}, and may shed light on the origin of FRBs and the emission mechanisms of neutron star mergers. This localisation could plausibly be improved by using distance constraints from the 3D gravitational wave skymap and estimating the distance of the coherent emission from its dispersion measure \citep[DM;][]{2003ApJ...598L..79I,2004MNRAS.348..999I}, although current DM-distance estimates are quite inaccurate due to large uncertainties in the DM contribution from the Milky Way Halo \citep{2015MNRAS.451.4277D} and the host galaxy. Therefore while we can narrow the search for the host galaxy we are unlikely to be able to uniquely identify it solely from prompt emission detected in fly's-eye mode.

There is also the prospect for an archival search for gravitational wave events coincident with ASKAP detections of FRBs \citep{2016PhRvD..93l2008A}.\\

\section{An optimised pointing strategy}
\subsection{Observing gravitational-wave source localisation regions}
\label{sec:observing_regions}
The first four observed gravitational wave events were detected by the LIGO Hanford and LIGO Livingston detectors and localised to areas ranging between 520 and 1600\,$\deg^2$ \citep{2017PhRvL.118v1101A,2017ApJ...851L..35A}. The first three-detector detection (using Advanced LIGO and Advanced Virgo), GW170814, was localised to 60\,$\deg^2$ \citep{2017PhRvL.119n1101A}, while GW170817 was localised to 28\,$\deg^2$ \citep{2017PhRvL.119p1101A}. During O3 the expected median 90\% localisation area is 120--180\,$\deg^2$, with up to 20\% of events having localisation areas smaller than 20\,$\deg^2$ \citep{2018LRR....21....3A}. Future observing runs, with improved sensitivity in existing detectors and the addition of the KAGRA and LIGO-India detectors, may be able to localise the majority of events to areas ${<12\,\deg^2}$ \citep{2018LRR....21....3A}.

It is not always feasible to observe entire localisation regions due to the large amount of observing time required, and therefore an optimised observing strategy can significantly increase the probability of successfully detecting an electromagnetic counterpart and decrease the observing time required. Observing a smaller fraction of the localisation area can significantly reduce the search area. For networks of three or four gravitational wave detectors, observing the 50\% confidence region reduces the search area by up to a factor of ten compared with the 90\% confidence region \citep{2011PhRvD..83j2001K,2014MNRAS.443..738B}. Using simulated skymaps from \citet{2014ApJ...795..105S} we find that the 50\% confidence region for two and three detector networks is on average five times, and up to twenty times, smaller than the 90\% confidence region. 

 \cite{2017ApJ...846...62S} used search strategies based on models of afterglow lightcurves and other parameters to determine spatial and temporal observing strategies that improve observing efficiencies. Other optimisation strategies \citep[e.g.][]{2016ExA....42..165C,2017ApJ...848L..33A} use estimates of afterglow luminosity to determine how time should be allocated between candidate hosts. However, current models of radio emission from neutron star mergers (see Section \ref{sec:longterm_emission}) are much less well constrained than kilonova models, and are also highly dependent on the circum-merger environment, making this approach less compelling for radio follow-up.

A strategy that has been implemented in follow-up using telescopes with a small ($\sim$ arcminute) FoV is targeted observations of specific galaxies in the localisation volume \citep[e.g.][]{2016MNRAS.462.1591E,2017ApJ...848L..33A}. We discuss applying a galaxy targeted technique for ASKAP in Section \ref{sec:galaxy_targeting}.

\begin{figure}
    \animategraphics[width=\linewidth,controls]{1.2}{greedy_animation_frames/}{1}{14}
    \caption{Illustration of the greedy ranked tiling strategy. Red lines correspond to 50\% (solid) and 90\% (dashed) localisation contours for an illustrative skymap. The gravitational wave skymap is covered with a pre-defined, overlapping grid of tiles (grey), ranked by their total enclosed probability. The highest ranked tile (blue) is selected and the probability in the region enclosed by it is set to zero. The tiles are then re-ranked and the next tile chosen, until the desired probability target is reached.}
    \label{fig:greedy_animation}
\end{figure}

\subsection{Tiling strategies for wide-field telescopes}
\citet{2016A&A...592A..82G} discuss possible pointing strategies for wide-field optical telescopes, and consider each pointing as a discrete, pre-determined tile placed on the sky. Gravitational wave sky-maps contain the three-dimensional (Right Ascension, Declination, distance) probability distribution function (PDF) of the event using the Hierarchical Equal Area isoLatitude Pixelization (HEALPix\footnote{\url{http://healpix.sourceforge.net/}}) projection, which divides the sky into pixels of equal surface area. The coverage of a single tile is then the sum of the PDF across all pixels contained in it.

The sensitivity of a radio telescope is not uniform across the FoV. Traditional radio interferometers with single pixel feeds have primary beams with highest sensitivity at the beam centre, and decreasing radially within the central lobe. An observing strategy using rectangular tiles will therefore result in non-uniform sensitivity across the localisation region. In contrast, the $6\times6$ square arrangement of beams in the ASKAP PAF produces a FoV that is accurately represented by a square tile. The sensitivity across a single pointing is not uniform, but uniform sensitivity across the interior beams can be achieved by interleaving multiple pointings. We represent the ASKAP FoV as a $5.5\times5.5$\,$\deg$ and allow a $0.5\,\deg$ overlap between tiles to ensure uniform sensitivity around the edges. 

In this section we outline four possible tiling strategies that could be used for ASKAP observations.

\subsubsection{Contour Covering tiles}
The simplest strategy to cover a gravitational wave localisation region is to cover entirely a given probability contour, for example, 90\%. We consider the minimum area containing 90\% of the event PDF (the 90\% contour) and select all tiles from a pre-computed grid that contains any part of the contour, to achieve at least 90\% coverage. This strategy is sub-optimal because it will cover significant extraneous area if the shorter axis of the localisation ellipse is the width of a few tiles or smaller.

\subsubsection{Ranked tiles}
A more effective way to minimise the extraneous area observed is the method of ranked tiles. Each tile is ranked by its sky coverage and tiles are chosen such that the desired confidence level is achieved with the minimum number of tiles. For this paper we implement both of the above strategies adapted for the ASKAP FoV as defined in \citet{2016A&A...592A..82G}.

\subsubsection{Greedy ranked tiles}
\label{sec:greedy_ranked_tiles}
While the approach above worked well for a single set of non-overlapping tiles, newer facilities have the ability to have multiple sets of fixed tiles, or even to not use fixed sets of tiles at all. Therefore \citet{2017PASP..129k4503G} introduced the method of greedy ranked tiles, which builds upon the ranked tiles strategy and allows overlapping tiles to be used without double-counting the probability contained in the overlapping region. The probability in the highest ranked tile is added to the cumulative probability coverage and then the area covered by the tile is zeroed out on the skymap. The process is then repeated until the desired overall coverage is achieved. An animation of this process is shown in Figure \ref{fig:greedy_animation}. We use the same strategy defined by \citet{2017PASP..129k4503G}, adapted for  the ASKAP FoV.

\begin{figure}
    \animategraphics[width=\linewidth,controls]{1.2}{shifted_animation_frames/}{1}{19}
    \caption{Illustration of the shifted ranked tiling strategy. Red lines correspond to 50\% (solid) and 90\% (dashed) localisation contours for an illustrative skymap. The gravitational wave skymap is covered with a single pre-defined grid of non-overlapping tiles (light grey). The ranked tiles required to cover the desired probability level are selected (dark grey) and grouped into strips of constant Declination. Each group is iteratively shifted in Right Ascension and the shift resulting in the desired probability enclosed within the minimum number of tiles. The set of optimally shifted tiles are then ranked and any extraneous tiles removed, leaving the optimal set of tiles for the entire localisation (blue).}
    \label{fig:shifted_animation}
\end{figure}

\subsubsection{Shifted ranked tiles}
\label{sec:shifted_ranked_tiles}
\citet{2016A&A...592A..82G} proposed that the ranked tiles strategy could be further optimised by using a grid of tiles free to be shifted in strips of constant declination. In this paper our strategy differs slightly as we only shift groups of adjacent tiles together, rather than entire strips of declination. Our implementation of the shifted ranked tiles strategy is defined as follows:

\begin{enumerate}
	\item Calculate the ranked tiles given by the primary sky grid and select the tiles that cover a given confidence region (for this example we use 90\%).
	
	\item Find the tiles that are adjacent to those selected above and include those in our sample. This gives the ranked tiles with an additional one-tile buffer, which we call the test-set of tiles. All other tiles are discarded for computational efficiency.
	
    \item Split the test-set into groups of constant declination. For each declination strip, group adjacent tiles together. This is required as some gravitational wave skymaps are multimodal (e.g. bimodal, multiple lobes, crosses) and shifting entire strips of declination for these events is suboptimal.
    
    \item For each group of adjacent tiles, shift the tiles in increments of Right Ascension. Re-calculate the ranked tiles for the test-set and record the number of tiles required for at least 90\% coverage and the actual coverage percentage at each shift. The optimal shift is then the shift which produces the required coverage in the least number of tiles. If multiple shifts produce the required coverage with the same number of tiles, choose the shift that maximises the coverage percentage. Repeat for each strip of declination in the test-set.
	
	\item Apply the optimal shift in each declination strip to the test-set and re-calculate the ranked tiles. 
\end{enumerate}
Because the ranked tile grid is periodic, optimisation only needs to be performed for one tile-width. An animation of this process is shown in Figure \ref{fig:shifted_animation}.

\subsection{An observing strategy for ASKAP}
Both the greedy and shifted ranked tile strategies are more complex and require more computation time compared to the other two strategies. The added burden of these strategies may outweigh any benefits they provide for some telescopes, but not for others, and here we consider an observing strategy designed specifically for ASKAP. Three main factors dictate the way we search for gravitational wave afterglows with ASKAP: our wide FoV, our transient detection strategy and our large observational latency. In optimising our observing strategy, we must consider all three factors.

\subsubsection{Field of View}
ASKAP has a FoV of $30\,\deg^{2}$ --- significantly larger than all other gigahertz-frequency radio telescopes and larger than many optical telescopes involved in gravitational wave follow-up \citep{2017ApJ...848L..12A}. The sky localisation capability of a telescope does not scale linearly with the FoV, as telescopes with a FoV comparable to the dimensions of the localisation region cover significantly more area that is extraneous to the localisation contour. The wide FoV of ASKAP allows localisation regions to be covered with fewer pointings than other telescopes, meaning that improving our observing strategy by a small number of pointings can result in significantly larger fractional improvements.

\subsubsection{Detection of transients and variables}
Transients surveys with optical telescopes typically use image differencing, where a deep reference image is subtracted from the observed image and removes all objects that are constant with time. The difficulty in image subtraction lies in matching parameters, including the instrument calibration and point-spread function, between the two science and observed images. This is easier if both images cover the same region of sky, so surveys for optical transients gain significant advantage from using a pre-defined grid of pointings.

However, in radio astronomy image differencing is rarely used due to the incomplete $u-v$ coverage of interferometers which causes the point-spread function (the dirty beam) to be more complex than a simple Airy disc. This introduces image artefacts that can easily be mistaken for transient sources in a difference image. Transient surveys with ASKAP use the Variables and Slow Transients pipeline \citep[VAST;][]{2013PASA...30....6M}, which detects transients via a process of source-finding, cross-matching and lightcurve analysis, meaning there is currently no advantage in using a pre-defined grid of pointings. Improved transient detection methods such as using image subtraction for LST-aligned observations are being tested with ASKAP (Stewart et al. in prep.).

\subsubsection{Latency}
Calculating an optimal pointing strategy can be time consuming. \citet{2017PASP..129k4503G} find that calculating the shifted ranked tiles strategy for some skymaps can take up to an hour, which is significant when searching for optical/ultraviolet/infrared emission from a counterpart, which may peak within hours of the merger. For any prompt follow-up we require a strategy that can be computed on timescales of seconds.

The time taken to compute the pointing strategy is not a concern when searching for long-term radio emission, which may not peak until tens--hundreds of days post-merger. Additionally, the current process of detecting gravitational wave events has multiple stages of localisation and parameter estimation, with each stage being more accurate but slower to compute. Currently an event can be detected in seconds, rapidly localised in minutes and full parameters calculated between hours and days post-merger \citep{2014ApJ...795..105S}. For GW170817 the event was initially localised to a 90\% credible region of 31\,$\deg^{2}$ four hours post-merger \citep{GCN21513}, which was refined to 28\,$\deg^{2}$ two months post-merger \citep{GCN21983} and finally 16\,$\deg^{2}$ nine months post-merger \citep{2019PhRvX...9a1001A}.

When searching for long-term emission, the lack of latency constraints allows us to wait for the improved parameter estimates, which will typically result in a smaller localisation region and therefore less required observing time. The localisation region can also change significantly once the full parameter estimate has been computed, as it also includes further noise removal and better detector calibration. For example, the full parameter localisation of GW170814 was 40\% smaller than the rapid localisation with only a small overlap between the two \citep[Fig. 3 of][]{2017PhRvL.119n1101A}. This advantage is not applicable to searches for prompt radio emission.

\subsection{Galaxy targeting}
\label{sec:galaxy_targeting}
The observing time required for follow-up of an event can be reduced by targeting galaxies within the localisation volume. This technique was employed in the follow-up of GW170817 where the localisation region was initially narrowed to 54 candidate host galaxies \citep[e.g.][]{GCN21519} before the optical counterpart was detected in the third ranked galaxy, NGC 4993.

We use the Galaxy List for the Advanced Detector Era \citep[GLADE,][]{2018MNRAS.479.2374D}, an all-sky galaxy catalogue aimed at supporting electromagnetic follow-up of gravitational wave events. The catalogue contains 3,262,883 objects, although measurements of blue luminosity are only available for half of them. GLADE is complete to a distance of 37 Mpc and is 54\% complete to the minimum Advanced LIGO O3 detector horizon of 120 Mpc \citep{2018MNRAS.479.2374D}.

A single ASKAP pointing can contain hundreds of galaxies. Averaging across the sky, excluding the Galactic Plane ($|b|<10\deg$), we find a typical ASKAP pointing contains 169 GLADE galaxies within a distance of 200\,Mpc and with a blue luminosity greater than $10^9$\,$L_{B,\odot}$. GLADE is $\sim 40\%$ complete at 200\,Mpc, so we expect a typical ASKAP pointing to contain 420 galaxies subject to these constraints. Targeting individual galaxies is therefore inefficient and instead we focus observations on regions containing a high density of candidate host galaxies by convolving the gravitational wave skymap with a galaxy catalogue using the method described by \citet{2016MNRAS.462.1591E}. For each pixel, $P_{\rm GW}(D)$ is the probability that the event occurred in that pixel at a distance $D$. This is calculated assuming a Gaussian distribution and the mean and standard deviation given by the gravitational wave skymap.

We determine the completeness of the galaxy catalogue as a function of distance, $C(D)$, as in \cite{2011CQGra..28h5016W}, from the $B$-luminosity for each galaxy. We then compare the cumulative blue luminosity of galaxies in the catalogue with the expected value, using the blue luminosity density derived from the SDSS survey by \citet{2008ApJ...675.1459K}, $(1.98 \pm 0.16)\times 10^{-2}L_{10}\textrm{ Mpc}^{-3}$ (where $L_{10}=10^{10}L_{B,\odot}$) as a reference. Within $\sim 35$ Mpc the calculated completeness exceeds 100\% (i.e. the expected value is an underestimate) and in this case we set $C(D)=1$.

We now consider each galaxy contained in the pixel and calculate $P_{\rm g}(D)$, the total probability that there is a galaxy at a given distance, $D$, weighted by its blue luminosity. The GLADE catalogue does not include distance uncertainty for most galaxies, and where this is the case we assume a 10\% uncertainty. The combined probability for each pixel is then given by
\begin{align}
	\label{eq:convolved_probability}
	P_p = P_{\rm GW}(D)P_{\rm g}(D)C(D) + P_{\rm GW}(D)(1-C(D)),
\end{align}
where the first factor corresponds to the event occurring in a catalogued galaxy, and the second corresponds to the event occurring in a non-catalogued galaxy. The convolved skymap is then normalised such that $\sum_{\rm all sky} P_p = 1$, so that the tiling efficiency can be easily compared between simulated events.

The catalogue of simulated events that we employ for testing our algorithms uses randomly generated event positions distributed homogeneously across the sky. To test the galaxy-targeting strategy we use the same method as \citet{2016MNRAS.462.1591E} and first consider whether or not the event occurred in a galaxy in our catalogue. We generate a random number, $0<x<1$, and compare it with the completeness of the galaxy catalogue at the distance of the event. If $x\geq C(D)$ then the event is considered to have occurred in an uncatalogued galaxy and we simply convolve the skymap and catalogue using equation \ref{eq:convolved_probability}. If $x<C(D)$, the event is treated as having occurred in a catalogued galaxy. We randomly choose a galaxy from the catalogue, weighted by $P_{\rm g}(D)$, the probability that the galaxy is located at the distance of the merger. We then rotate the positions of every galaxy in the catalogue so that the position of the selected host galaxy is consistent with the location of the simulated event.

\section{Comparison of pointing strategies}
\subsection{Coverage of simulated skymaps}
We tested our pointing strategy using 475 skymaps from the 2016 O2 scenario described in \citet{2014ApJ...795..105S}. This consists of the two Advanced LIGO detectors and Advanced Virgo operating on an independent and random (i.e. downtime is not correlated between detectors) 80\% duty cycle, with a neutron star merger range of 108 Mpc for the two Advanced LIGO detectors and 36 Mpc for Advanced Virgo. This is greater than the achieved sensitivity during O2, but below the expected O3 sensitivity of 120--170 Mpc for Advanced LIGO and 65--85 Mpc for Advanced Virgo \citep{2018LRR....21....3A}.

We consider all 475 skymaps and compare the outcome when covering them with contour-covering, simple ranked tiles, greedy ranked tiles or the shifted ranked tile approach. We have considered probability coverage targets ranging from 20\% to 95\%, as approaching 100\% coverage drastically increases computation time while producing diminishing returns. The contour-covering and greedy ranked tiles methods typically take seconds to compute for a single skymap, while the shifted ranked method takes minutes. Computing the reusable set of tiles for both ranked tiles strategies has an additional overhead of 10--20 hours depending on the desired resolution. We evaluated each method based on the average probability density per tile it achieves, and the fraction of events it detects. We do not consider whether the radio emission from the event would have been detectable, as the models for possible emission vary by orders of magnitude in temporal behaviour and peak flux density, and instead consider an event to be detected if it occurred within a tile.

We use the ranked and contour-covering tiles strategies as a baseline for comparing the shifted and greedy ranked tiles methods, both for the original gravitational wave skymaps and the convolved galaxy-targeted skymaps. The galaxy targeting strategy was then tested using the greedy optimisation strategy.

\begin{figure}
	\includegraphics[width=\columnwidth,trim={0cm 0cm 0cm 0cm},clip]{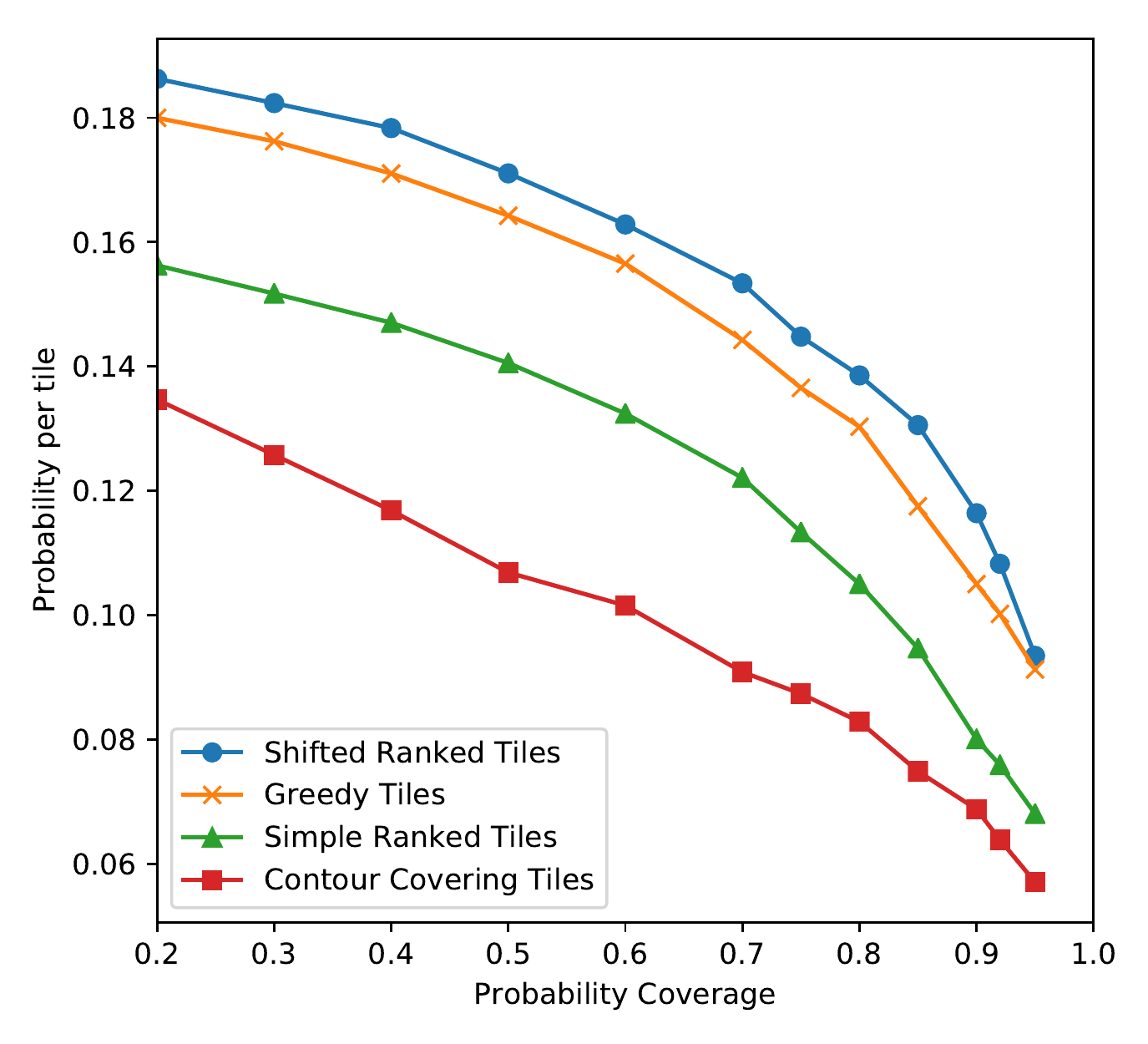}
	\caption{Covered probability per tile, for all four tiling strategies averaged across their application to 475 simulated skymaps. As expected, the two optimised strategies significantly outperform the contour covering and simple ranked tiled strategies.}
	\label{fig:coverage_density}
\end{figure}

\subsection{Comparing advanced ranked tiling strategies}
Figure \ref{fig:coverage_density} shows the average probability coverage per tile as a function of target probability for all four tiling strategies using the gravitational wave skymaps. As expected, the contour-covering strategy is the least efficient, while the greedy and shifted ranked tile strategies outperform the simple ranked tile strategy by 2.4 and 3.2 percentage points per tile on average. At 90\% probability coverage this corresponds to the shifted ranked tile strategy producing $\sim$10\% more efficient coverage per tile compared to the greedy ranked tile strategy.

\begin{figure}
	\includegraphics[width=\columnwidth,trim={0cm 0cm 0cm 0cm},clip]{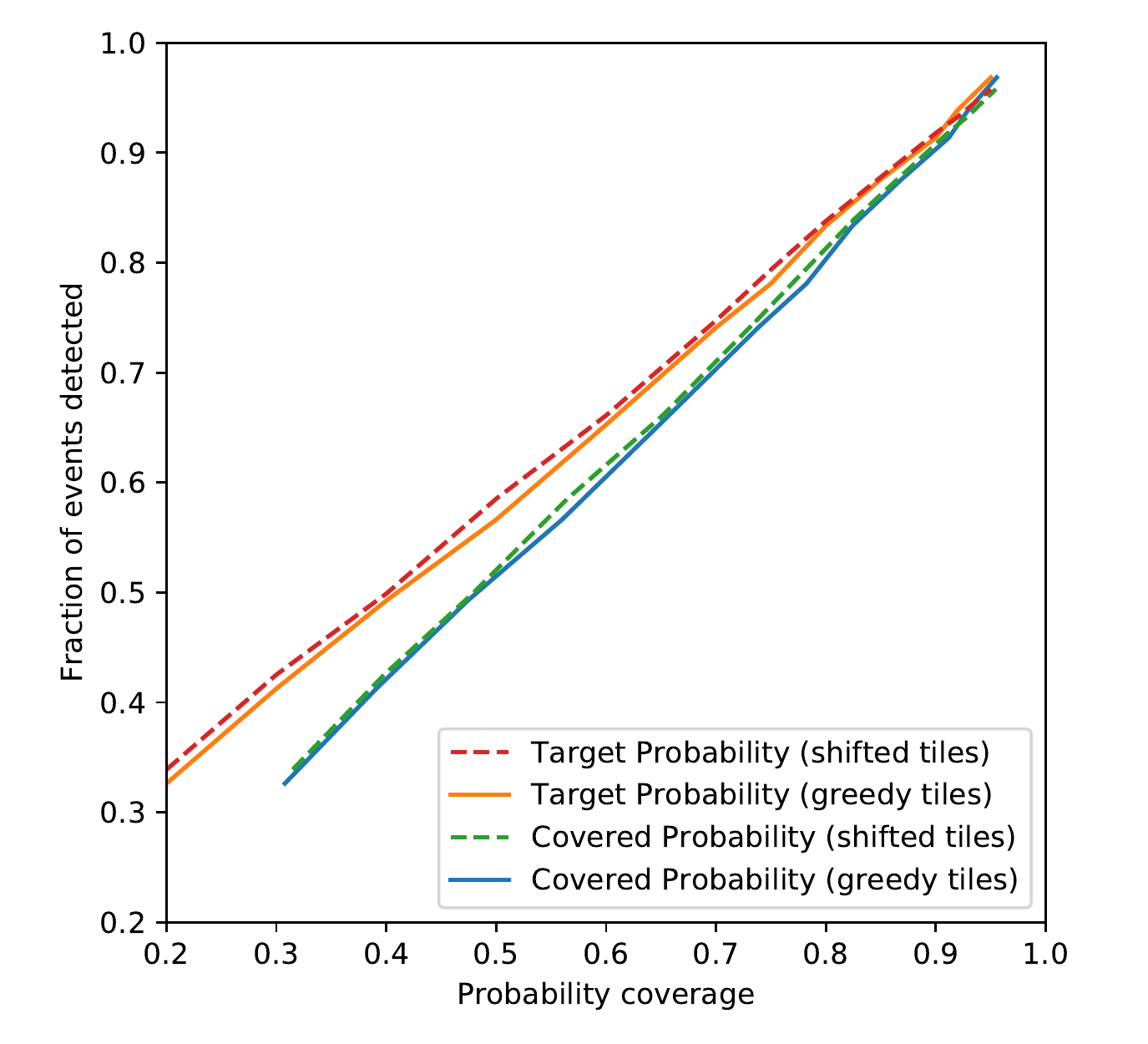}
	\caption{Fraction of events detected  as a function of target probability coverage (orange/red) and actual probability coverage (blue/green) for both the greedy tiles (solid) and shifted tiles (dashed) methods. The actual probability coverage is calculated by taking the mean coverage for all simulations.}
	\label{fig:detection_frac}
\end{figure}

We also consider our ability to detect events. Figure \ref{fig:detection_frac} shows the fraction of all 475 events that were detected as a function of probability coverage for both the shifted and greedy tiles. Due to the wide FoV of ASKAP, even an optimal tiling strategy results in extraneous coverage, so we have plotted the detection fraction as a function of both the target and achieved probability coverage averaged across all events. Both tiling strategies produce similar results across the entire probability range. As expected, the detection fraction is approximately equal to the covered probability, confirming that Advanced LIGO-Virgo skymaps do accurately recreate the probability distribution function of the event.

As expected, the shifted ranked tiles strategy slightly outperforms the greedy ranked tiles strategy. For most skymaps they will achieve identical results, but in some cases the greedy strategy produces a sub-optimal strategy as it only considers the next-most optimal tile at each step rather than the final overall tiling. This can result in gaps between tiles, or cases where more tiles are used than necessary. For example, if there is a strip of probability 2 tiles wide the greedy strategy will place the first tile at the centre of the strip, resulting in 3 tiles being required to cover the localisation (e.g. compare the tiling in Figures \ref{fig:greedy_animation} and \ref{fig:shifted_animation}). The shifted ranked tiles strategy does not have these issues, as it considers the placement of independent groups of adjacent tiles. Each iteration computes the total probability coverage of the final overally tiling, and by considering adjacent tiles there are no gaps.

The actual gain in telescope time is dependent on the localisation area, the desired probability coverage, the total number of observing epochs and the time allocated per epoch. However, if we consider a probability coverage of 90\% and an allocation of 1 hour per pointing we can estimate the gain in telescope time for specific events. Simulated event 493048 is detected with the LIGO-Hanford and Advanced Virgo detectors and has a 90\% localisation of $135\,\deg^2$, comparable the median detector horizon for O3. The simple ranked tiles strategy requires 15 tiles to cover the event, while the shifted and greedy ranked tiles strategies require 10 and 11 tiles respectively, corresponding to savings of 5 hours per epoch. The total savings increase with the time per pointing and the total number of epochs.

As a result, the shifted ranked tiles strategy should be used when the required observing latency is longer than the typical computation time (on the order of minutes). For low-latency observations the greedy ranked tiles strategy is preferable due to its shorter computation time.

\begin{figure}
	\includegraphics[width=\columnwidth,trim={0cm 0cm 0cm 0cm},clip]{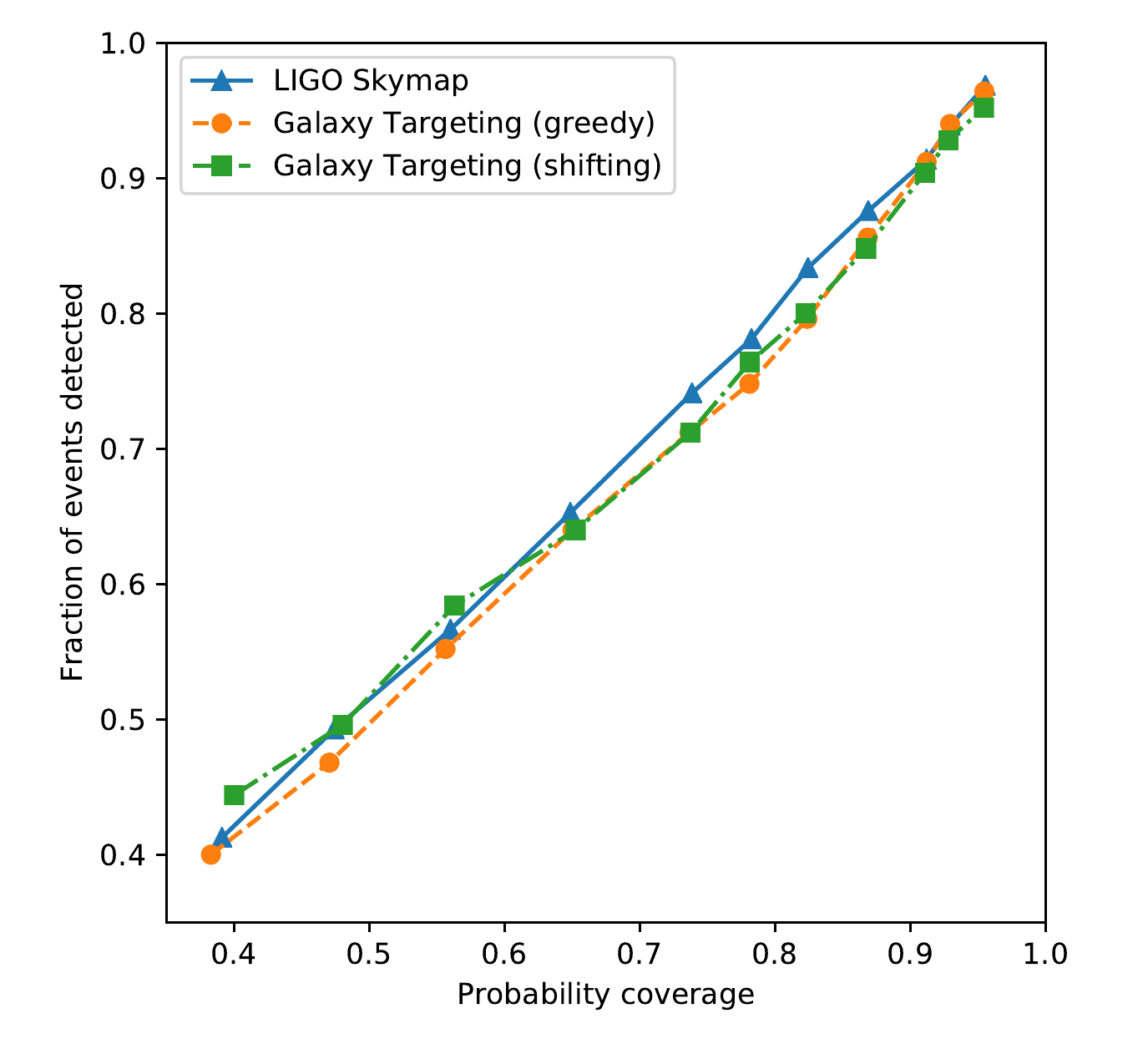}
	\caption{Comparison of sky coverage using the galaxy targeted probability map and the original gravitational wave skymap.}
	\label{fig:galaxy_targeting}
\end{figure}

\subsection{Galaxy targeting strategy}
We applied the greedy and shifted ranked tile strategies to all 250 simulated skymaps that have 3D localisations. Figure \ref{fig:galaxy_targeting} shows the fraction of events covered by both strategies when applied to the galaxy targeted skymap, and the greedy strategy applied to the original skymap, as a function of probability coverage of the original skymap. There is no appreciable difference between the two strategies, although for mid-range probability coverages, original skymap marginally outperforms the galaxy targeted skymap, and for higher probability coverages ($>0.9$) the galaxy targeted skymap performs marginally better. We propose that this negligible difference is the result of the incompleteness of the galaxy catalogue, as well as the wide ASKAP FoV that is more suitable for surveying large areas of the sky than targeting individual sources. For most events the optimal tiling strategy produced was nearly identical for the two different skymaps. We expect that the effectiveness of using a galaxy targeting strategy for most telescopes will significantly improve in the near future, as wider and deeper galaxy surveys produce more complete catalogues. However, it is unclear whether this improvement will affect ASKAP.

Conversely, we note that improvements to gravitational wave detectors will result in larger detector horizons. Advanced LIGO is expected to detect neutron star mergers at a distance of 190 Mpc \citep{2018LRR....21....3A}, while next generation detectors like Cosmic Explorer may detect neutron star mergers at $z=6$ \citep{2017CQGra..34d4001A}. Distant events will have larger localisation volumes along with less-complete catalogues of potential host galaxies, making a galaxy targeting approach less feasible.

\section{Detectability of events with ASKAP}
\subsection{Sensitivity and angular resolution}
Table \ref{tab:askap_specs} shows the expected timeline and specifications as ASKAP moves towards design sensitivity. ASKAP should reach design specifications by the start of O3 in early 2019 and will be capable of observing entire localisation regions to sub-mJy levels in hours, or achieving a 12 $\mu$Jy image RMS in a 12 hour pointing.

While GW170817 would have been detectable with ASKAP 40--300 days post-merger (Figure \ref{fig:gw170817_prediction}), we emphasise that this event may not be typical of the population of binary neutron star systems expected in the future. The observed luminosity of the afterglow increases with the isotropic equivalent energy, and the circum-merger density, $n$, the velocity of the outflow and the fraction of the internal energy deposited into the outflowing electrons and the accompanying magnetic field. These parameters also determine the temporal evolution of the afterglow, although the exact scaling factors are model-dependant and not necessarily trivial to compute \citep[e.g.][]{1998ApJ...497L..17S,2011Natur.478...82N,2017MNRAS.471.1652L,2018MNRAS.473..576G,2018MNRAS.481.2581L}. The expected radio emission is also anisotropic, and highly dependent on the merger inclination angle, with on-axis events typically more luminous. The horizon of gravitational wave detectors is also larger for on-axis events, known as the gravitational wave Malmquist effect \citep{2011CQGra..28l5023S}.

Typical values for most of the source parameters are unclear, although it is reasonable to assume that inclination angle is uniformly distributed. The circum-merger density of GW170817 is estimated as $n\sim 10^{-4}$ cm$^{-3}$, significantly lower than the median circum-burst density of short GRBs ($3-15\times 10^{-3}$ cm$^{-3}$), with densities $n\sim 1$ cm$^{-3}$ not uncommon \citep{2015ApJ...815..102F}. However, we expect approximately half of all neutron star mergers to have circum-merger densities $n>0.1$ cm$^{-3}$ \citep{2016ApJ...831..190H}. Even at the limit of the Advanced LIGO detection range, these events may still be radio loud, peaking at a flux density of tens of mJy (Dobie et al. in prep.), which is readily detected in a short ASKAP observation.

The 10\,arcsec angular resolution of ASKAP at 1.4\,GHz is large compared to the $1-2$\,arcsec achievable with the Australia Telescope Compact Array (ATCA) or the Jansky Very Large Array (JVLA), and we will not always be able to resolve the GW counterpart from its host galaxy. Short GRBs have physical offsets between 0.5--75\,kpc from the host galaxy centre \citep[median 5\,kpc;][]{2014ARA&A..52...43B}, which \citet{2016ApJ...831..190H} found corresponds to 70\% of radio counterparts at 200\,Mpc contaminated by the galaxy core at ASKAP's angular resolution. In comparison, GW170817 was offset from the centre of NGC 4993 by 10.6\,arcsec \citep[or 2 kpc;][]{2017Sci...358.1556C}, which is at the limit of resolution with ASKAP at design specifications. Events close to the host nucleus will still be detectable with ASKAP if the radio counterpart has a peak flux density greater than that of the nucleus, which may be the the case for a majority of events \citep{2016ApJ...831..190H}.

Once an event has been detected with ASKAP, we expect the bulk of the monitoring campaign will be performed with the ATCA, the JVLA and potentially MeerKAT, due to their superior sensitivity and angular resolution. This will enable the lightcurve to be much more tightly constrained than is possible with ASKAP alone.

\subsection{Detection of false-positives}
A major concern in the search for optical counterparts to gravitational wave events is the detection of false positives. \citet{2013ApJ...767..124N} found that there could be tens to thousands of optical false positive detections per gravitational wave event. The rate of radio transients, on the other hand, is comparatively low, and the chance of a false-positive detection in the form of an unrelated radio transient is minimal. \citet{2011MNRAS.412..634B} found one transient at 843 MHz in a study spanning timescales of 1 day to 20 years, corresponding to an areal transient rate of $7.5\times 10^{-4}$\,$\deg^{-2}$ for flux densities >14\,mJy. Other works found the 1.4\,GHz radio transient areal density to be <0.1\,$\deg^{-2}$ \citep[>1\,mJy on timescales of 0.5 to 7 years;][]{2016MNRAS.461.3314H}, <0.37\,$\deg^{-2}$ \citep[>0.2\,mJy on timescales of 1 week to 3 months;][]{2013ApJ...768..165M} and <0.01\,$\deg^{-2}$ \citep[>1.5\,mJy on daily timescales;][]{2018MNRAS.478.1784B}.

False positives may also be detected as a result of an insufficiently sensitive reference image, although ultimately a reference image will be available from the Evolutionary Map of the Universe \citep[EMU,][]{2011PASA...28..215N}, an ASKAP survey of the sky south of $+30$\,$\deg$ to an image RMS of $\sim10\,\mu$Jy at 1.3 GHz, along with our own observations soon after the event. Until EMU is completed we will rely on the NRAO VLA Sky Survey \citep[NVSS,][]{1998AJ....115.1693C} which covers the sky north of $-40\,\deg$ with an image RMS of $\sim 0.45$\,mJy at 1.4 GHz, the VLA Sky Survey (VLASS\footnote{\url{https://science.nrao.edu/science/surveys/vlass}}) which will cover the sky north of $-40\,\deg$ to an image RMS of 67 $\mu$Jy at 2-4 GHz, or the Sydney University Molonglo Sky Survey \citep[SUMSS,][]{1999AJ....117.1578B,2003MNRAS.342.1117M} which covers the sky south of $-30\,\deg$ to an image RMS of $\sim 1$--$2.5$\,mJy at 843 MHz. The age of these surveys may also produce false positives, with long-term AGN variability caused by refractive interstellar scintillation being detected as transient behaviour on short timescales.

\subsection{Expectations for LIGO-Virgo's third observing run.}
The O3 observing run is expected to begin in April 2019, with a median detector horizon of 120-170 Mpc (Advanced Virgo: 65-85 Mpc) for neutron star mergers \citep{2018LRR....21....3A}. Given the current best estimate of the neutron star merger rate, $1.54^{+3.2}_{-1.22}\textrm{ Mpc}^{-3}\textrm{Myr}^{-1}$ \citep{2017PhRvL.119p1101A}, we expect there to be 1--50 significant neutron star merger triggers during O3. ASKAP is capable of observing the sky south of $+30$\,$\deg$, corresponding to 75\% of the whole sky. We expect to be able to observe a similar fraction of gravitational-wave triggers, although the real fraction may be lower given the geographically dependent sensitivity of gravitational wave detectors, any possible anisotropy in the distribution of neutron star mergers, and other detector constraints.

The localisation capability of Advanced LIGO for O3 and beyond is discussed in Section \ref{sec:observing_regions}, with the median 90\% probability containment approximately 150\,$\deg^2$ for O3, which is observable with a handful of ASKAP pointings.

The distance to GW170817 (40 Mpc) is significantly less than the expected O3 detector horizon, and was also well localised (90\% containment within $16\,\deg^{2}$) which allowed electromagnetic emission from the event to be detected by telescopes targeting candidate host galaxies. For events with a larger localisation volume, the number of candidate host galaxies may be large enough that they cannot all be targeted in a reasonable time. ASKAP will follow-up all observable neutron star merger events, with a focus on those poorly localised events with no detected electromagnetic counterpart.

\section{Conclusions}
The first detection of a neutron star merger, GW170817, was initially localised to $16\,\deg^{2}$ through its gravitational wave signal. This enabled an optical counterpart to be identified, and the event localised to a galaxy, within hours. However, future events are likely to have much larger localisation volumes (hundreds of $\deg^{2}$) and detecting the electromagnetic counterparts to these events will require wide-field telescopes. Once ASKAP reaches design specifications we expect it to play an important role in the follow-up of poorly localised events (particularly where no other electromagnetic counterpart has been detected) due to its wide FoV and high survey speed. Importantly, we will be able to use ASKAP to localise events that may not be observable at other wavelengths due to factors including solar angle and dust obscuration.

We have discussed prospects for ASKAP detecting two forms of radio emission from neutron star mergers (prompt coherent emission that may be similar to an FRB, and long-term emission similar to a standard sGRB afterglow or GW170817). We have compared four different methods of tiling localisation regions with telescope pointings, building on previous work using widefield optical telescopes. We find that our implementation of the shifted ranked tiles method (see Section \ref{sec:shifted_ranked_tiles}) outperforms previously investigated tiling methods for long-term ASKAP follow-up, while the greedy ranked tiles method \citep[see Section \ref{sec:greedy_ranked_tiles} \&][]{2017PASP..129k4503G} is preferable for prompt follow-up due to its faster computation time. We also find that there is no significant advantage to using a galaxy-targeting approach.

Applying these pointing strategies to follow-up of gravitational wave events will optimise the use of telescope time and maximise the chance of a detection, localising the event and allowing us to better understand its properties.

\begin{acknowledgements}
DD acknowledges useful discussions with H. Qiu. We thank the referee for helpful comments and suggestions which improved this paper.
DD is supported by an Australian Government Research Training Program Scholarship. TM acknowledges the support of the Australian Research Council through grant FT150100099. DLK was supported by NSF grant AST-1816492. SG was supported by NSF Award PHY-1607585. This research has made use of NASA's Astrophysics Data System Bibliographic Services.
The Australian SKA Pathfinder is part of the Australia Telescope National Facility which is funded by the Commonwealth of Australia for operation as a National Facility managed by CSIRO. The Murchison Radio-astronomy Observatory is managed by the CSIRO, who also provide operational support to ASKAP. We acknowledge the Wajarri Yamatji people as the traditional owners of the Observatory site.

\textit{Software:} Astropy \citep{2018AJ....156..123A}, Matplotlib \citep{2007CSE.....9...90H}, Numpy \citep{2011CSE....13b..22V}
\end{acknowledgements}

\bibliographystyle{pasa-mnras}
%\bibliography{bibliography}

\begin{thebibliography}{}
\makeatletter
\relax
\def\mn@urlcharsother{\let\do\@makeother \do\$\do\&\do\#\do\^\do\_\do\%\do\~}
\definecolor{darkblue}{rgb}{0,0,0.597656}
\def\mndoi{\begingroup\mn@urlcharsother \@ifnextchar [ {\mndoi@} {\mndoi@[]}}
\def\mndoi@[#1]#2{\def\@tempa{#1}\ifx\@tempa\@empty \href
  {http://dx.doi.org/#2} {\textcolor{darkblue}{doi:#2}}\else \href
  {http://dx.doi.org/#2} {\textcolor{darkblue}{#1}}\fi \endgroup}
\def\mn@eprint#1#2{\mn@eprint@#1:#2::\@nil}
\def\mn@eprint@arXiv#1{\href {http://arxiv.org/abs/#1} {{\tt arXiv:#1}}}
\def\mn@eprint@dblp#1{\href {http://dblp.uni-trier.de/rec/bibtex/#1.xml}
  {dblp:#1}}
\def\mn@eprint@#1:#2:#3:#4\@nil{\def\@tempa {#1}\def\@tempb {#2}\def\@tempc
  {#3}\ifx \@tempc \@empty \let \@tempc \@tempb \let \@tempb \@tempa \fi \ifx
  \@tempb \@empty \def\@tempb {arXiv}\fi \@ifundefined
  {mn@eprint@\@tempb}{\@tempb:\@tempc}{\expandafter \expandafter \csname
  mn@eprint@\@tempb\endcsname \expandafter{\@tempc}}}

\bibitem[\protect\citeauthoryear{{Abbott} et~al.,}{{Abbott}
  et~al.}{2016}]{2016PhRvD..93l2008A}
{Abbott} B.~P.,  et~al., 2016, \mndoi [\prd] {10.1103/PhysRevD.93.122008},
  \href {http://adsabs.harvard.edu/abs/2016PhRvD..93l2008A} {93, 122008}

\bibitem[\protect\citeauthoryear{{Abbott}, {Abbott}, {Abbott}, {Abernathy},
  {Ackley}  \& et al.}{{Abbott} et~al.}{2017a}]{2017CQGra..34d4001A}
{Abbott} B.~P.,  {Abbott} R.,  {Abbott} T.~D.,  {Abernathy} M.~R.,  {Ackley}
  K.,   et al. 2017a, \mndoi [Classical and Quantum Gravity]
  {10.1088/1361-6382/aa51f4}, \href
  {https://ui.adsabs.harvard.edu/#abs/2017CQGra..34d4001A} {34, 044001}

\bibitem[\protect\citeauthoryear{{Abbott}, {Abbott}, {Abbott}, {Acernese},
  {Ackley}  \& et al.}{{Abbott} et~al.}{2017b}]{2017PhRvL.118v1101A}
{Abbott} B.~P.,  {Abbott} R.,  {Abbott} T.~D.,  {Acernese} F.,  {Ackley} K.,
  et al. 2017b, \mndoi [\prl] {10.1103/PhysRevLett.118.221101}, \href
  {https://ui.adsabs.harvard.edu/#abs/2017PhRvL.118v1101A} {118, 221101}

\bibitem[\protect\citeauthoryear{{Abbott} et~al.,}{{Abbott}
  et~al.}{2017c}]{2017PhRvL.119n1101A}
{Abbott} B.~P.,  et~al., 2017c, \mndoi [Physical Review Letters]
  {10.1103/PhysRevLett.119.141101}, \href
  {http://adsabs.harvard.edu/abs/2017PhRvL.119n1101A} {119, 141101}

\bibitem[\protect\citeauthoryear{{Abbott} et~al.,}{{Abbott}
  et~al.}{2017d}]{2017PhRvL.119p1101A}
{Abbott} B.~P.,  et~al., 2017d, \mndoi [Physical Review Letters]
  {10.1103/PhysRevLett.119.161101}, \href
  {http://adsabs.harvard.edu/abs/2017PhRvL.119p1101A} {119, 161101}

\bibitem[\protect\citeauthoryear{{Abbott}, {Abbott}, {Abbott}, {Acernese},
  {Ackley}  \& et al.}{{Abbott} et~al.}{2017e}]{2017Natur.551...85A}
{Abbott} B.~P.,  {Abbott} R.,  {Abbott} T.~D.,  {Acernese} F.,  {Ackley} K.,
  et al. 2017e, \mndoi [\nat] {10.1038/nature24471}, \href
  {https://ui.adsabs.harvard.edu/#abs/2017Natur.551...85A} {551, 85}

\bibitem[\protect\citeauthoryear{{Abbott} et~al.,}{{Abbott}
  et~al.}{2017f}]{2017ApJ...848L..12A}
{Abbott} B.~P.,  et~al., 2017f, \mndoi [\apjl] {10.3847/2041-8213/aa91c9},
  \href {http://adsabs.harvard.edu/abs/2017ApJ...848L..12A} {848, L12}

\bibitem[\protect\citeauthoryear{{Abbott} et~al.,}{{Abbott}
  et~al.}{2017g}]{2017ApJ...848L..13A}
{Abbott} B.~P.,  et~al., 2017g, \mndoi [\apjl] {10.3847/2041-8213/aa920c},
  \href {http://adsabs.harvard.edu/abs/2017ApJ...848L..13A} {848, L13}

\bibitem[\protect\citeauthoryear{{Abbott}, {Abbott}, {Abbott}, {Acernese}  \&
  {Ackley}}{{Abbott} et~al.}{2017h}]{2017ApJ...851L..35A}
{Abbott} B.~P.,  {Abbott} R.,  {Abbott} T.~D.,  {Acernese} F.,   {Ackley} K.,
  2017h, \mndoi [\apj] {10.3847/2041-8213/aa9f0c}, \href
  {https://ui.adsabs.harvard.edu/#abs/2017ApJ...851L..35A} {851, L35}

\bibitem[\protect\citeauthoryear{{Abbott} et~al.,}{{Abbott}
  et~al.}{2018}]{2018LRR....21....3A}
{Abbott} B.~P.,  et~al., 2018, \mndoi [Living Reviews in Relativity]
  {10.1007/s41114-018-0012-9}, \href
  {http://adsabs.harvard.edu/abs/2018LRR....21....3A} {21, 3}

\bibitem[\protect\citeauthoryear{{Abbott}, {Abbott}, {Abbott}, {Acernese},
  {Ackley}  \& et al.}{{Abbott} et~al.}{2019}]{2019PhRvX...9a1001A}
{Abbott} B.~P.,  {Abbott} R.,  {Abbott} T.~D.,  {Acernese} F.,  {Ackley} K.,
  et al. 2019, \mndoi [Physical Review X] {10.1103/PhysRevX.9.011001}, \href
  {https://ui.adsabs.harvard.edu/\#abs/2019PhRvX...9a1001A} {9, 011001}

\bibitem[\protect\citeauthoryear{{Alexander} et~al.,}{{Alexander}
  et~al.}{2018}]{2018ApJ...863L..18A}
{Alexander} K.~D.,  et~al., 2018, \mndoi [\apj] {10.3847/2041-8213/aad637},
  \href {https://ui.adsabs.harvard.edu/#abs/2018ApJ...863L..18A} {863, L18}

\bibitem[\protect\citeauthoryear{{Andreoni} et~al.,}{{Andreoni}
  et~al.}{2017}]{2017PASA...34...69A}
{Andreoni} I.,  et~al., 2017, \mndoi [\pasa] {10.1017/pasa.2017.65}, \href
  {http://adsabs.harvard.edu/abs/2017PASA...34...69A} {34, e069}

\bibitem[\protect\citeauthoryear{{Arcavi} et~al.,}{{Arcavi}
  et~al.}{2017}]{2017ApJ...848L..33A}
{Arcavi} I.,  et~al., 2017, \mndoi [\apjl] {10.3847/2041-8213/aa910f}, \href
  {http://adsabs.harvard.edu/abs/2017ApJ...848L..33A} {848, L33}

\bibitem[\protect\citeauthoryear{{Aso}, {Michimura}, {Somiya}, {Ando},
  {Miyakawa}, {Sekiguchi}, {Tatsumi}  \& {Yamamoto}}{{Aso}
  et~al.}{2013}]{2013PhRvD..88d3007A}
{Aso} Y.,  {Michimura} Y.,  {Somiya} K.,  {Ando} M.,  {Miyakawa} O.,
  {Sekiguchi} T.,  {Tatsumi} D.,   {Yamamoto} H.,  2013, \mndoi [\prd]
  {10.1103/PhysRevD.88.043007}, \href
  {https://ui.adsabs.harvard.edu/\#abs/2013PhRvD..88d3007A} {88, 043007}

\bibitem[\protect\citeauthoryear{{Astropy Collaboration} et~al.,}{{Astropy
  Collaboration} et~al.}{2018}]{2018AJ....156..123A}
{Astropy Collaboration} et~al., 2018, \mndoi [\aj] {10.3847/1538-3881/aabc4f},
  \href {https://ui.adsabs.harvard.edu/#abs/2018AJ....156..123A} {156, 123}

\bibitem[\protect\citeauthoryear{{Bannister}, {Murphy}, {Gaensler}, {Hunstead}
  \& {Chatterjee}}{{Bannister} et~al.}{2011}]{2011MNRAS.412..634B}
{Bannister} K.~W.,  {Murphy} T.,  {Gaensler} B.~M.,  {Hunstead} R.~W.,
  {Chatterjee} S.,  2011, \mndoi [\mnras] {10.1111/j.1365-2966.2010.17938.x},
  \href {http://adsabs.harvard.edu/abs/2011MNRAS.412..634B} {412, 634}

\bibitem[\protect\citeauthoryear{{Bannister} et~al.,}{{Bannister}
  et~al.}{2017a}]{2017ApJ...841L..12B}
{Bannister} K.~W.,  et~al., 2017a, \mndoi [\apjl] {10.3847/2041-8213/aa71ff},
  \href {http://adsabs.harvard.edu/abs/2017ApJ...841L..12B} {841, L12}

\bibitem[\protect\citeauthoryear{{Bannister}, {Shannon}, {Hotan}, {James},
  {Macquart}, {Oslowski}  \& {Farah}}{{Bannister} et~al.}{2017b}]{GCN21562}
{Bannister} K.,  {Shannon} R.,  {Hotan} A.,  {James} C.,  {Macquart} J.-P.,
  {Oslowski} S.,   {Farah} W.,  2017b, GCN, 21562

\bibitem[\protect\citeauthoryear{{Bannister}, {Shannon}, {Hotan}, {James},
  {Macquart}, {Oslowski}  \& {Farah}}{{Bannister} et~al.}{2017c}]{GCN21671}
{Bannister} K.,  {Shannon} R.,  {Hotan} A.,  {James} C.,  {Macquart} J.-P.,
  {Oslowski} S.,   {Farah} W.,  2017c, GCN, 21671

\bibitem[\protect\citeauthoryear{{Bartos} et~al.,}{{Bartos}
  et~al.}{2014}]{2014MNRAS.443..738B}
{Bartos} I.,  et~al., 2014, \mndoi [\mnras] {10.1093/mnras/stu1205}, \href
  {http://adsabs.harvard.edu/abs/2014MNRAS.443..738B} {443, 738}

\bibitem[\protect\citeauthoryear{{Berger}}{{Berger}}{2014}]{2014ARA&A..52...43B}
{Berger} E.,  2014, \mndoi [Annual Review of Astronomy and Astrophysics]
  {10.1146/annurev-astro-081913-035926}, \href
  {https://ui.adsabs.harvard.edu/#abs/2014ARA&A..52...43B} {52, 43}

\bibitem[\protect\citeauthoryear{{Bhandari} et~al.,}{{Bhandari}
  et~al.}{2018}]{2018MNRAS.478.1784B}
{Bhandari} S.,  et~al., 2018, \mndoi [\mnras] {10.1093/mnras/sty1157}, \href
  {https://ui.adsabs.harvard.edu/#abs/2018MNRAS.478.1784B} {478, 1784}

\bibitem[\protect\citeauthoryear{{Bock}, {Large}  \& {Sadler}}{{Bock}
  et~al.}{1999}]{1999AJ....117.1578B}
{Bock} D.~C.-J.,  {Large} M.~I.,   {Sadler} E.~M.,  1999, \mndoi [\aj]
  {10.1086/300786}, \href {http://adsabs.harvard.edu/abs/1999AJ....117.1578B}
  {117, 1578}

\bibitem[\protect\citeauthoryear{{Chan}, {Messenger}, {Heng}  \&
  {Hendry}}{{Chan} et~al.}{2018}]{2018PhRvD..97l3014C}
{Chan} M.~L.,  {Messenger} C.,  {Heng} I.~S.,   {Hendry} M.,  2018, \mndoi
  [\prd] {10.1103/PhysRevD.97.123014}, \href
  {https://ui.adsabs.harvard.edu/#abs/2018PhRvD..97l3014C} {97, 123014}

\bibitem[\protect\citeauthoryear{{Chu}, {Howell}, {Rowlinson}, {Gao}, {Zhang},
  {Tingay}, {Bo{\"e}r}  \& {Wen}}{{Chu} et~al.}{2016}]{2016MNRAS.459..121C}
{Chu} Q.,  {Howell} E.~J.,  {Rowlinson} A.,  {Gao} H.,  {Zhang} B.,  {Tingay}
  S.~J.,  {Bo{\"e}r} M.,   {Wen} L.,  2016, \mndoi [\mnras]
  {10.1093/mnras/stw576}, \href
  {https://ui.adsabs.harvard.edu/#abs/2016MNRAS.459..121C} {459, 121}

\bibitem[\protect\citeauthoryear{{Condon}, {Cotton}, {Greisen}, {Yin},
  {Perley}, {Taylor}  \& {Broderick}}{{Condon}
  et~al.}{1998}]{1998AJ....115.1693C}
{Condon} J.~J.,  {Cotton} W.~D.,  {Greisen} E.~W.,  {Yin} Q.~F.,  {Perley}
  R.~A.,  {Taylor} G.~B.,   {Broderick} J.~J.,  1998, \mndoi [\aj]
  {10.1086/300337}, \href {http://adsabs.harvard.edu/abs/1998AJ....115.1693C}
  {115, 1693}

\bibitem[\protect\citeauthoryear{{Cook}, {Van Sistine}, {Singer}  \&
  {Kasliwal}}{{Cook} et~al.}{2017}]{GCN21519}
{Cook} D.~O.,  {Van Sistine} A.,  {Singer} L.,   {Kasliwal} M.~M.,  2017, GCN,
  21519

\bibitem[\protect\citeauthoryear{{Corsi} et~al.,}{{Corsi}
  et~al.}{2018}]{2018ApJ...861L..10C}
{Corsi} A.,  et~al., 2018, \mndoi [\apj] {10.3847/2041-8213/aacdfd}, \href
  {https://ui.adsabs.harvard.edu/#abs/2018ApJ...861L..10C} {861, L10}

\bibitem[\protect\citeauthoryear{{Coughlin} \& {Stubbs}}{{Coughlin} \&
  {Stubbs}}{2016}]{2016ExA....42..165C}
{Coughlin} M.,  {Stubbs} C.,  2016, \mndoi [Experimental Astronomy]
  {10.1007/s10686-016-9503-4}, \href
  {https://ui.adsabs.harvard.edu/\#abs/2016ExA....42..165C} {42, 165}

\bibitem[\protect\citeauthoryear{{Coulter} et~al.,}{{Coulter}
  et~al.}{2017}]{2017Sci...358.1556C}
{Coulter} D.~A.,  et~al., 2017, \mndoi [Science] {10.1126/science.aap9811},
  \href {http://adsabs.harvard.edu/abs/2017Sci...358.1556C} {358, 1556}

\bibitem[\protect\citeauthoryear{{D'Avanzo} et~al.,}{{D'Avanzo}
  et~al.}{2018}]{2018A&A...613L...1D}
{D'Avanzo} P.,  et~al., 2018, \mndoi [\aap] {10.1051/0004-6361/201832664},
  \href {https://ui.adsabs.harvard.edu/#abs/2018A&A...613L...1D} {613, L1}

\bibitem[\protect\citeauthoryear{{D{\'a}lya} et~al.,}{{D{\'a}lya}
  et~al.}{2018}]{2018MNRAS.479.2374D}
{D{\'a}lya} G.,  et~al., 2018, \mndoi [\mnras] {10.1093/mnras/sty1703}, \href
  {https://ui.adsabs.harvard.edu/#abs/2018MNRAS.479.2374D} {479, 2374}

\bibitem[\protect\citeauthoryear{{Dobie}, {Hotan}, {Bannister}, {Kaplan},
  {Murphy}  \& {Lynch}}{{Dobie} et~al.}{2017a}]{GCN21625}
{Dobie} D.,  {Hotan} A.,  {Bannister} K.,  {Kaplan} D.,  {Murphy} T.,   {Lynch}
  C.,  2017a, GCN, 21625

\bibitem[\protect\citeauthoryear{{Dobie}, {Hotan}, {Bannister}, {Kaplan},
  {Murphy}  \& {Lynch}}{{Dobie} et~al.}{2017b}]{GCN21639}
{Dobie} D.,  {Hotan} A.,  {Bannister} K.,  {Kaplan} D.,  {Murphy} T.,   {Lynch}
  C.,  2017b, GCN, 21639

\bibitem[\protect\citeauthoryear{{Dobie} et~al.,}{{Dobie}
  et~al.}{2018}]{2018ApJ...858L..15D}
{Dobie} D.,  et~al., 2018, \mndoi [\apjl] {10.3847/2041-8213/aac105}, \href
  {http://adsabs.harvard.edu/abs/2018ApJ...858L..15D} {858, L15}

\bibitem[\protect\citeauthoryear{{Dolag}, {Gaensler}, {Beck}  \&
  {Beck}}{{Dolag} et~al.}{2015}]{2015MNRAS.451.4277D}
{Dolag} K.,  {Gaensler} B.~M.,  {Beck} A.~M.,   {Beck} M.~C.,  2015, \mndoi
  [\mnras] {10.1093/mnras/stv1190}, \href
  {https://ui.adsabs.harvard.edu/#abs/2015MNRAS.451.4277D} {451, 4277}

\bibitem[\protect\citeauthoryear{{Evans} et~al.,}{{Evans}
  et~al.}{2016}]{2016MNRAS.462.1591E}
{Evans} P.~A.,  et~al., 2016, \mndoi [\mnras] {10.1093/mnras/stw1746}, \href
  {http://adsabs.harvard.edu/abs/2016MNRAS.462.1591E} {462, 1591}

\bibitem[\protect\citeauthoryear{{Evans} et~al.,}{{Evans}
  et~al.}{2017}]{2017Sci...358.1565E}
{Evans} P.~A.,  et~al., 2017, \mndoi [Science] {10.1126/science.aap9580}, \href
  {https://ui.adsabs.harvard.edu/#abs/2017Sci...358.1565E} {358, 1565}

\bibitem[\protect\citeauthoryear{{Falcke} \& {Rezzolla}}{{Falcke} \&
  {Rezzolla}}{2014}]{2014A&A...562A.137F}
{Falcke} H.,  {Rezzolla} L.,  2014, \mndoi [\aap]
  {10.1051/0004-6361/201321996}, \href
  {http://adsabs.harvard.edu/abs/2014A%26A...562A.137F} {562, A137}

\bibitem[\protect\citeauthoryear{{Fong}, {Berger}, {Margutti}  \&
  {Zauderer}}{{Fong} et~al.}{2015}]{2015ApJ...815..102F}
{Fong} W.,  {Berger} E.,  {Margutti} R.,   {Zauderer} B.~A.,  2015, \mndoi
  [\apj] {10.1088/0004-637X/815/2/102}, \href
  {http://adsabs.harvard.edu/abs/2015ApJ...815..102F} {815, 102}

\bibitem[\protect\citeauthoryear{{Ghosh}, {Bloemen}, {Nelemans}, {Groot}  \&
  {Price}}{{Ghosh} et~al.}{2016}]{2016A&A...592A..82G}
{Ghosh} S.,  {Bloemen} S.,  {Nelemans} G.,  {Groot} P.~J.,   {Price} L.~R.,
  2016, \mndoi [\aap] {10.1051/0004-6361/201527712}, \href
  {http://adsabs.harvard.edu/abs/2016A%26A...592A..82G} {592, A82}

\bibitem[\protect\citeauthoryear{{Ghosh}, {Chatterjee}, {Kaplan}, {Brady}  \&
  {Van Sistine}}{{Ghosh} et~al.}{2017}]{2017PASP..129k4503G}
{Ghosh} S.,  {Chatterjee} D.,  {Kaplan} D.~L.,  {Brady} P.~R.,   {Van Sistine}
  A.,  2017, \mndoi [Publications of the Astronomical Society of the Pacific]
  {10.1088/1538-3873/aa884f}, \href
  {https://ui.adsabs.harvard.edu/#abs/2017PASP..129k4503G} {129, 114503}

\bibitem[\protect\citeauthoryear{{Goldstein} et~al.,}{{Goldstein}
  et~al.}{2017}]{2017ApJ...848L..14G}
{Goldstein} A.,  et~al., 2017, \mndoi [\apjl] {10.3847/2041-8213/aa8f41}, \href
  {http://adsabs.harvard.edu/abs/2017ApJ...848L..14G} {848, L14}

\bibitem[\protect\citeauthoryear{{Gottlieb}, {Nakar}  \& {Piran}}{{Gottlieb}
  et~al.}{2018}]{2018MNRAS.473..576G}
{Gottlieb} O.,  {Nakar} E.,   {Piran} T.,  2018, \mndoi [\mnras]
  {10.1093/mnras/stx2357}, \href
  {http://adsabs.harvard.edu/abs/2018MNRAS.473..576G} {473, 576}

\bibitem[\protect\citeauthoryear{{Granot} et~al.}{{Granot}
  et~al.}{2002}]{2002ApJ...570L..61G}
{Granot} J.,  et~al., 2002, \mndoi [\apjl] {10.1086/340991}, \href
  {http://adsabs.harvard.edu/abs/2002ApJ...570L..61G} {570, L61}

\bibitem[\protect\citeauthoryear{{Haggard}, {Nynka}, {Ruan}, {Kalogera},
  {Cenko}, {Evans}  \& {Kennea}}{{Haggard} et~al.}{2017}]{2017ApJ...848L..25H}
{Haggard} D.,  {Nynka} M.,  {Ruan} J.~J.,  {Kalogera} V.,  {Cenko} S.~B.,
  {Evans} P.,   {Kennea} J.~A.,  2017, \mndoi [\apjl]
  {10.3847/2041-8213/aa8ede}, \href
  {http://adsabs.harvard.edu/abs/2017ApJ...848L..25H} {848, L25}

\bibitem[\protect\citeauthoryear{{Hallinan} et~al.,}{{Hallinan}
  et~al.}{2017}]{2017Sci...358.1579H}
{Hallinan} G.,  et~al., 2017, \mndoi [Science] {10.1126/science.aap9855}, \href
  {http://adsabs.harvard.edu/abs/2017Sci...358.1579H} {358, 1579}

\bibitem[\protect\citeauthoryear{Hampson et~al.,}{Hampson
  et~al.}{2012}]{6328742}
Hampson G.,  et~al., 2012, in 2012 International Conference on Electromagnetics
  in Advanced Applications. pp 807--809, \mndoi{10.1109/ICEAA.2012.6328742}

\bibitem[\protect\citeauthoryear{{Hancock}, {Drury}, {Bell}, {Murphy}  \&
  {Gaensler}}{{Hancock} et~al.}{2016}]{2016MNRAS.461.3314H}
{Hancock} P.~J.,  {Drury} J.~A.,  {Bell} M.~E.,  {Murphy} T.,   {Gaensler}
  B.~M.,  2016, \mndoi [\mnras] {10.1093/mnras/stw1486}, \href
  {http://adsabs.harvard.edu/abs/2016MNRAS.461.3314H} {461, 3314}

\bibitem[\protect\citeauthoryear{{Hotokezaka} \& {Piran}}{{Hotokezaka} \&
  {Piran}}{2015}]{2015MNRAS.450.1430H}
{Hotokezaka} K.,  {Piran} T.,  2015, \mndoi [\mnras] {10.1093/mnras/stv620},
  \href {http://adsabs.harvard.edu/abs/2015MNRAS.450.1430H} {450, 1430}

\bibitem[\protect\citeauthoryear{{Hotokezaka}, {Nissanke}, {Hallinan}, {Lazio},
  {Nakar}  \& {Piran}}{{Hotokezaka} et~al.}{2016}]{2016ApJ...831..190H}
{Hotokezaka} K.,  {Nissanke} S.,  {Hallinan} G.,  {Lazio} T.~J.~W.,  {Nakar}
  E.,   {Piran} T.,  2016, \mndoi [\apj] {10.3847/0004-637X/831/2/190}, \href
  {http://adsabs.harvard.edu/abs/2016ApJ...831..190H} {831, 190}

\bibitem[\protect\citeauthoryear{{Hotokezaka}, {Nakar}, {Gottlieb}, {Nissanke},
  {Masuda}, {Hallinan}, {Mooley}  \& {Deller}}{{Hotokezaka}
  et~al.}{2018}]{2018arXiv180610596H}
{Hotokezaka} K.,  {Nakar} E.,  {Gottlieb} O.,  {Nissanke} S.,  {Masuda} K.,
  {Hallinan} G.,  {Mooley} K.~P.,   {Deller} A.~T.,  2018, preprint, \href
  {https://ui.adsabs.harvard.edu/#abs/2018arXiv180610596H} {p.
  arXiv:1806.10596} (\mn@eprint {arXiv} {1806.10596})

\bibitem[\protect\citeauthoryear{{Hunter}}{{Hunter}}{2007}]{2007CSE.....9...90H}
{Hunter} J.~D.,  2007, \mndoi [Computing in Science and Engineering]
  {10.1109/MCSE.2007.55}, \href
  {https://ui.adsabs.harvard.edu/#abs/2007CSE.....9...90H} {9, 90}

\bibitem[\protect\citeauthoryear{{Inoue}}{{Inoue}}{2004}]{2004MNRAS.348..999I}
{Inoue} S.,  2004, \mndoi [\mnras] {10.1111/j.1365-2966.2004.07359.x}, \href
  {https://ui.adsabs.harvard.edu/#abs/2004MNRAS.348..999I} {348, 999}

\bibitem[\protect\citeauthoryear{{Ioka}}{{Ioka}}{2003}]{2003ApJ...598L..79I}
{Ioka} K.,  2003, \mndoi [\apj] {10.1086/380598}, \href
  {https://ui.adsabs.harvard.edu/#abs/2003ApJ...598L..79I} {598, L79}

\bibitem[\protect\citeauthoryear{{Johnston} et~al.,}{{Johnston}
  et~al.}{2008}]{2008ExA....22..151J}
{Johnston} S.,  et~al., 2008, \mndoi [Experimental Astronomy]
  {10.1007/s10686-008-9124-7}, \href
  {http://adsabs.harvard.edu/abs/2008ExA....22..151J} {22, 151}

\bibitem[\protect\citeauthoryear{{Klimenko} et~al.,}{{Klimenko}
  et~al.}{2011}]{2011PhRvD..83j2001K}
{Klimenko} S.,  et~al., 2011, \mndoi [\prd] {10.1103/PhysRevD.83.102001}, \href
  {http://adsabs.harvard.edu/abs/2011PhRvD..83j2001K} {83, 102001}

\bibitem[\protect\citeauthoryear{{Kopparapu}, {Hanna}, {Kalogera},
  {O'Shaughnessy}, {Gonz{\'a}lez}, {Brady}  \& {Fairhurst}}{{Kopparapu}
  et~al.}{2008}]{2008ApJ...675.1459K}
{Kopparapu} R.~K.,  {Hanna} C.,  {Kalogera} V.,  {O'Shaughnessy} R.,
  {Gonz{\'a}lez} G.,  {Brady} P.~R.,   {Fairhurst} S.,  2008, \mndoi [\apj]
  {10.1086/527348}, \href {http://adsabs.harvard.edu/abs/2008ApJ...675.1459K}
  {675, 1459}

\bibitem[\protect\citeauthoryear{{Lamb}, {Mandel}  \& {Resmi}}{{Lamb}
  et~al.}{2018}]{2018MNRAS.481.2581L}
{Lamb} G.~P.,  {Mandel} I.,   {Resmi} L.,  2018, \mndoi [\mnras]
  {10.1093/mnras/sty2196}, \href
  {https://ui.adsabs.harvard.edu/\#abs/2018MNRAS.481.2581L} {481, 2581}

\bibitem[\protect\citeauthoryear{{Lazzati}, {Deich}, {Morsony}  \&
  {Workman}}{{Lazzati} et~al.}{2017}]{2017MNRAS.471.1652L}
{Lazzati} D.,  {Deich} A.,  {Morsony} B.~J.,   {Workman} J.~C.,  2017, \mndoi
  [\mnras] {10.1093/mnras/stx1683}, \href
  {https://ui.adsabs.harvard.edu/\#abs/2017MNRAS.471.1652L} {471, 1652}

\bibitem[\protect\citeauthoryear{{Lazzati}, {Perna}, {Morsony}, {Lopez-Camara},
  {Cantiello}, {Ciolfi}, {Giacomazzo}  \& {Workman}}{{Lazzati}
  et~al.}{2018}]{2018PhRvL.120x1103L}
{Lazzati} D.,  {Perna} R.,  {Morsony} B.~J.,  {Lopez-Camara} D.,  {Cantiello}
  M.,  {Ciolfi} R.,  {Giacomazzo} B.,   {Workman} J.~C.,  2018, \mndoi [\prl]
  {10.1103/PhysRevLett.120.241103}, \href
  {https://ui.adsabs.harvard.edu/#abs/2018PhRvL.120x1103L} {120, 241103}

\bibitem[\protect\citeauthoryear{{Lipunov} \& {Panchenko}}{{Lipunov} \&
  {Panchenko}}{1996}]{1996A&A...312..937L}
{Lipunov} V.~M.,  {Panchenko} I.~E.,  1996, \aap, \href
  {https://ui.adsabs.harvard.edu/\#abs/1996A&A...312..937L} {312, 937}

\bibitem[\protect\citeauthoryear{{Lorimer}, {Bailes}, {McLaughlin}, {Narkevic}
  \& {Crawford}}{{Lorimer} et~al.}{2007}]{2007Sci...318..777L}
{Lorimer} D.~R.,  {Bailes} M.,  {McLaughlin} M.~A.,  {Narkevic} D.~J.,
  {Crawford} F.,  2007, \mndoi [Science] {10.1126/science.1147532}, \href
  {http://adsabs.harvard.edu/abs/2007Sci...318..777L} {318, 777}

\bibitem[\protect\citeauthoryear{{Macquart}, {Shannon}, {Bannister}, {James},
  {Ekers}  \& {Bunton}}{{Macquart} et~al.}{2019}]{2019ApJ...872L..19M}
{Macquart} J.~P.,  {Shannon} R.~M.,  {Bannister} K.~W.,  {James} C.~W.,
  {Ekers} R.~D.,   {Bunton} J.~D.,  2019, \mndoi [\apj]
  {10.3847/2041-8213/ab03d6}, \href
  {https://ui.adsabs.harvard.edu/\#abs/2019ApJ...872L..19M} {872, L19}

\bibitem[\protect\citeauthoryear{{Mahony} et~al.,}{{Mahony}
  et~al.}{2018}]{2018ApJ...867L..10M}
{Mahony} E.~K.,  et~al., 2018, \mndoi [\apj] {10.3847/2041-8213/aae7cb}, \href
  {https://ui.adsabs.harvard.edu/#abs/2018ApJ...867L..10M} {867, L10}

\bibitem[\protect\citeauthoryear{{Margutti} et~al.,}{{Margutti}
  et~al.}{2018}]{2018ApJ...856L..18M}
{Margutti} R.,  et~al., 2018, \mndoi [\apjl] {10.3847/2041-8213/aab2ad}, \href
  {http://adsabs.harvard.edu/abs/2018ApJ...856L..18M} {856, L18}

\bibitem[\protect\citeauthoryear{{Mauch}, {Murphy}, {Buttery}, {Curran},
  {Hunstead}, {Piestrzynski}, {Robertson}  \& {Sadler}}{{Mauch}
  et~al.}{2003}]{2003MNRAS.342.1117M}
{Mauch} T.,  {Murphy} T.,  {Buttery} H.~J.,  {Curran} J.,  {Hunstead} R.~W.,
  {Piestrzynski} B.,  {Robertson} J.~G.,   {Sadler} E.~M.,  2003, \mndoi
  [\mnras] {10.1046/j.1365-8711.2003.06605.x}, \href
  {http://adsabs.harvard.edu/abs/2003MNRAS.342.1117M} {342, 1117}

\bibitem[\protect\citeauthoryear{{Metzger} \& {Berger}}{{Metzger} \&
  {Berger}}{2012}]{2012ApJ...746...48M}
{Metzger} B.~D.,  {Berger} E.,  2012, \mndoi [\apj]
  {10.1088/0004-637X/746/1/48}, \href
  {http://adsabs.harvard.edu/abs/2012ApJ...746...48M} {746, 48}

\bibitem[\protect\citeauthoryear{{Mooley}, {Frail}, {Ofek}, {Miller},
  {Kulkarni}  \& {Horesh}}{{Mooley} et~al.}{2013}]{2013ApJ...768..165M}
{Mooley} K.~P.,  {Frail} D.~A.,  {Ofek} E.~O.,  {Miller} N.~A.,  {Kulkarni}
  S.~R.,   {Horesh} A.,  2013, \mndoi [\apj] {10.1088/0004-637X/768/2/165},
  \href {http://adsabs.harvard.edu/abs/2013ApJ...768..165M} {768, 165}

\bibitem[\protect\citeauthoryear{{Mooley} et~al.,}{{Mooley}
  et~al.}{2018a}]{2018Natur.554..207M}
{Mooley} K.~P.,  et~al., 2018a, \mndoi [\nat] {10.1038/nature25452}, \href
  {http://adsabs.harvard.edu/abs/2018Natur.554..207M} {554, 207}

\bibitem[\protect\citeauthoryear{{Mooley} et~al.,}{{Mooley}
  et~al.}{2018b}]{2018Natur.561..355M}
{Mooley} K.~P.,  et~al., 2018b, \mndoi [\nat] {10.1038/s41586-018-0486-3},
  \href {https://ui.adsabs.harvard.edu/#abs/2018Natur.561..355M} {561, 355}

\bibitem[\protect\citeauthoryear{{Mooley} et~al.,}{{Mooley}
  et~al.}{2018c}]{2018ApJ...868L..11M}
{Mooley} K.~P.,  et~al., 2018c, \mndoi [\apj] {10.3847/2041-8213/aaeda7}, \href
  {https://ui.adsabs.harvard.edu/#abs/2018ApJ...868L..11M} {868, L11}

\bibitem[\protect\citeauthoryear{{Murphy} et~al.,}{{Murphy}
  et~al.}{2013}]{2013PASA...30....6M}
{Murphy} T.,  et~al., 2013, \mndoi [\pasa] {10.1017/pasa.2012.006}, \href
  {http://adsabs.harvard.edu/abs/2013PASA...30....6M} {30, e006}

\bibitem[\protect\citeauthoryear{{Nakar} \& {Piran}}{{Nakar} \&
  {Piran}}{2011}]{2011Natur.478...82N}
{Nakar} E.,  {Piran} T.,  2011, \mndoi [\nat] {10.1038/nature10365}, \href
  {http://adsabs.harvard.edu/abs/2011Natur.478...82N} {478, 82}

\bibitem[\protect\citeauthoryear{{Nakar} \& {Piran}}{{Nakar} \&
  {Piran}}{2018}]{2018MNRAS.478..407N}
{Nakar} E.,  {Piran} T.,  2018, \mndoi [\mnras] {10.1093/mnras/sty952}, \href
  {https://ui.adsabs.harvard.edu/#abs/2018MNRAS.478..407N} {478, 407}

\bibitem[\protect\citeauthoryear{{Nissanke}, {Kasliwal}  \&
  {Georgieva}}{{Nissanke} et~al.}{2013}]{2013ApJ...767..124N}
{Nissanke} S.,  {Kasliwal} M.,   {Georgieva} A.,  2013, \mndoi [\apj]
  {10.1088/0004-637X/767/2/124}, \href
  {http://adsabs.harvard.edu/abs/2013ApJ...767..124N} {767, 124}

\bibitem[\protect\citeauthoryear{{Norris} et~al.,}{{Norris}
  et~al.}{2011}]{2011PASA...28..215N}
{Norris} R.~P.,  et~al., 2011, \mndoi [\pasa] {10.1071/AS11021}, \href
  {http://adsabs.harvard.edu/abs/2011PASA...28..215N} {28, 215}

\bibitem[\protect\citeauthoryear{{Petroff} et~al.,}{{Petroff}
  et~al.}{2016}]{2016PASA...33...45P}
{Petroff} E.,  et~al., 2016, \mndoi [\pasa] {10.1017/pasa.2016.35}, \href
  {http://adsabs.harvard.edu/abs/2016PASA...33...45P} {33, e045}

\bibitem[\protect\citeauthoryear{{Pshirkov} \& {Postnov}}{{Pshirkov} \&
  {Postnov}}{2010}]{2010Ap&SS.330...13P}
{Pshirkov} M.~S.,  {Postnov} K.~A.,  2010, \mndoi [\apss]
  {10.1007/s10509-010-0395-x}, \href
  {https://ui.adsabs.harvard.edu/\#abs/2010Ap&SS.330...13P} {330, 13}

\bibitem[\protect\citeauthoryear{{Resmi} et~al.,}{{Resmi}
  et~al.}{2018}]{2018ApJ...867...57R}
{Resmi} L.,  et~al., 2018, \mndoi [\apj] {10.3847/1538-4357/aae1a6}, \href
  {https://ui.adsabs.harvard.edu/\#abs/2018ApJ...867...57R} {867, 57}

\bibitem[\protect\citeauthoryear{{Salafia}, {Colpi}, {Branchesi},
  {Chassande-Mottin}, {Ghirlanda}, {Ghisellini}  \& {Vergani}}{{Salafia}
  et~al.}{2017}]{2017ApJ...846...62S}
{Salafia} O.~S.,  {Colpi} M.,  {Branchesi} M.,  {Chassande-Mottin} E.,
  {Ghirlanda} G.,  {Ghisellini} G.,   {Vergani} S.~D.,  2017, \mndoi [\apj]
  {10.3847/1538-4357/aa850e}, \href
  {http://adsabs.harvard.edu/abs/2017ApJ...846...62S} {846, 62}

\bibitem[\protect\citeauthoryear{{Sari}, {Piran}  \& {Narayan}}{{Sari}
  et~al.}{1998}]{1998ApJ...497L..17S}
{Sari} R.,  {Piran} T.,   {Narayan} R.,  1998, \mndoi [\apj] {10.1086/311269},
  \href {https://ui.adsabs.harvard.edu/\#abs/1998ApJ...497L..17S} {497, L17}

\bibitem[\protect\citeauthoryear{{Schutz}}{{Schutz}}{2011}]{2011CQGra..28l5023S}
{Schutz} B.~F.,  2011, \mndoi [Classical and Quantum Gravity]
  {10.1088/0264-9381/28/12/125023}, \href
  {https://ui.adsabs.harvard.edu/\#abs/2011CQGra..28l5023S} {28, 125023}

\bibitem[\protect\citeauthoryear{Shannon et~al.,}{Shannon
  et~al.}{2018}]{shannon2018dispersion}
Shannon R.,  et~al., 2018, Nature, p.~1

\bibitem[\protect\citeauthoryear{{Singer} et~al.,}{{Singer}
  et~al.}{2014}]{2014ApJ...795..105S}
{Singer} L.~P.,  et~al., 2014, \mndoi [\apj] {10.1088/0004-637X/795/2/105},
  \href {http://adsabs.harvard.edu/abs/2014ApJ...795..105S} {795, 105}

\bibitem[\protect\citeauthoryear{{Somiya}}{{Somiya}}{2012}]{2012CQGra..29l4007S}
{Somiya} K.,  2012, \mndoi [Classical and Quantum Gravity]
  {10.1088/0264-9381/29/12/124007}, \href
  {https://ui.adsabs.harvard.edu/\#abs/2012CQGra..29l4007S} {29, 124007}

\bibitem[\protect\citeauthoryear{{The LIGO Scientific Collaboration} \& {the
  Virgo Collaboration}}{{The LIGO Scientific Collaboration} \& {the Virgo
  Collaboration}}{2017a}]{GCN21513}
{The LIGO Scientific Collaboration} {the Virgo Collaboration} 2017a, GCN, 21513

\bibitem[\protect\citeauthoryear{{The LIGO Scientific Collaboration} \& {the
  Virgo Collaboration}}{{The LIGO Scientific Collaboration} \& {the Virgo
  Collaboration}}{2017b}]{GCN21983}
{The LIGO Scientific Collaboration} {the Virgo Collaboration} 2017b, GCN, 21983

\bibitem[\protect\citeauthoryear{{Thornton} et~al.,}{{Thornton}
  et~al.}{2013}]{2013Sci...341...53T}
{Thornton} D.,  et~al., 2013, \mndoi [Science] {10.1126/science.1236789}, \href
  {http://adsabs.harvard.edu/abs/2013Sci...341...53T} {341, 53}

\bibitem[\protect\citeauthoryear{{Totani}}{{Totani}}{2013}]{2013PASJ...65L..12T}
{Totani} T.,  2013, \mndoi [\pasj] {10.1093/pasj/65.5.L12}, \href
  {http://adsabs.harvard.edu/abs/2013PASJ...65L..12T} {65, L12}

\bibitem[\protect\citeauthoryear{{Troja} et~al.,}{{Troja}
  et~al.}{2017}]{2017Natur.551...71T}
{Troja} E.,  et~al., 2017, \mndoi [\nat] {10.1038/nature24290}, \href
  {http://adsabs.harvard.edu/abs/2017Natur.551...71T} {551, 71}

\bibitem[\protect\citeauthoryear{{Unnikrishnan}}{{Unnikrishnan}}{2013}]{2013IJMPD..2241010U}
{Unnikrishnan} C.~S.,  2013, \mndoi [International Journal of Modern Physics D]
  {10.1142/S0218271813410101}, \href
  {https://ui.adsabs.harvard.edu/\#abs/2013IJMPD..2241010U} {22, 1341010}

\bibitem[\protect\citeauthoryear{{Wang}, {Yang}, {Wu}, {Dai}  \& {Wang}}{{Wang}
  et~al.}{2016}]{2016ApJ...822L...7W}
{Wang} J.-S.,  {Yang} Y.-P.,  {Wu} X.-F.,  {Dai} Z.-G.,   {Wang} F.-Y.,  2016,
  \mndoi [\apjl] {10.3847/2041-8205/822/1/L7}, \href
  {http://adsabs.harvard.edu/abs/2016ApJ...822L...7W} {822, L7}

\bibitem[\protect\citeauthoryear{{White}, {Daw}  \& {Dhillon}}{{White}
  et~al.}{2011}]{2011CQGra..28h5016W}
{White} D.~J.,  {Daw} E.~J.,   {Dhillon} V.~S.,  2011, \mndoi [Classical and
  Quantum Gravity] {10.1088/0264-9381/28/8/085016}, \href
  {http://adsabs.harvard.edu/abs/2011CQGra..28h5016W} {28, 085016}

\bibitem[\protect\citeauthoryear{{Zhang}}{{Zhang}}{2014}]{2014ApJ...780L..21Z}
{Zhang} B.,  2014, \mndoi [\apjl] {10.1088/2041-8205/780/2/L21}, \href
  {http://adsabs.harvard.edu/abs/2014ApJ...780L..21Z} {780, L21}

\bibitem[\protect\citeauthoryear{{van Eerten}, {Ryan}, {Ricci}, {Burgess},
  {Wieringa}, {Piro}, {Cenko}  \& {Sakamoto}}{{van Eerten}
  et~al.}{2018}]{2018arXiv180806617V}
{van Eerten} E. T.~H.,  {Ryan} G.,  {Ricci} R.,  {Burgess} J.~M.,  {Wieringa}
  M.,  {Piro} L.,  {Cenko} S.~B.,   {Sakamoto} T.,  2018, preprint, \href
  {https://ui.adsabs.harvard.edu/#abs/2018arXiv180806617V} {p.
  arXiv:1808.06617} (\mn@eprint {arXiv} {1808.06617})

\bibitem[\protect\citeauthoryear{{van der Walt}, {Colbert}  \&
  {Varoquaux}}{{van der Walt} et~al.}{2011}]{2011CSE....13b..22V}
{van der Walt} S.,  {Colbert} S.~C.,   {Varoquaux} G.,  2011, \mndoi [Computing
  in Science and Engineering] {10.1109/MCSE.2011.37}, \href
  {https://ui.adsabs.harvard.edu/#abs/2011CSE....13b..22V} {13, 22}

\makeatother
\end{thebibliography}
\input{bibliography.bbl}

\end{document}